\documentclass[aps,twocolumn,showpacs,preprintnumbers,nofootinbib,prd,10pt,superscriptaddress,floats,floatfix]{revtex4-1}
	
\usepackage{subcaption}
\usepackage{ragged2e}
\DeclareCaptionJustification{justified}{\justifying}
\usepackage{amssymb,amsmath,verbatim,mathtools,needspace,enumitem,etoolbox,graphicx,physics,microtype,afterpage,bm}
\usepackage{esint}
\usepackage[dvipsnames]{xcolor}
\definecolor{linkcolor}{rgb}{0.0,0.3,0.5}
\usepackage[unicode, colorlinks=true, linkcolor=linkcolor, citecolor=linkcolor, filecolor=linkcolor,urlcolor=linkcolor, pdfusetitle]{hyperref}
\usepackage[all]{hypcap}
\usepackage[T1]{fontenc}
\usepackage[utf8]{inputenc}
\usepackage{tabularx}
\usepackage{float}
\newcommand\underrel[3][]{\mathrel{\mathop{#3}\limits_{%
      \ifx c#1\relax\mathclap{#2}\else#2\fi}}}
\usepackage{soul}
\usepackage{lmodern}
\allowdisplaybreaks
\usepackage{tikz}
\usepackage{cleveref}
\usepackage{color}
\usepackage{framed}
\usepackage{hyperref}
\hypersetup{colorlinks, citecolor=bluscuro, linkcolor=black, urlcolor=bluscuro}
\definecolor{rossos}{cmyk}{0,1,1,0.55}
\definecolor{bluscuro}{rgb}{0.15, 0.2, .85}

\newcommand{\be}{\begin{equation}}
\newcommand{\ee}{\end{equation}}
\renewcommand{\d}{{\mathrm d}}

\def\lsim{\mathrel{\rlap{\lower4pt\hbox{\hskip0.5pt$\sim$}}
    \raise1pt\hbox{$<$}}}         
\def\gsim{\mathrel{\rlap{\lower4pt\hbox{\hskip0.5pt$\sim$}}
    \raise1pt\hbox{$>$}}}         

\makeatletter
\newcommand{\subsetsim}{\mathrel{\mathpalette\subset@sim\relax}}
\newcommand{\subset@sim}[2]{%
  \vtop{\offinterlineskip\m@th
    \ialign{\hfil##\cr
     ~$#1\subset$\cr\noalign{\kern0.5pt}\scalebox{0.9}{$#1\sim$}\cr
    }%
  }%
}
\makeatother
\makeatletter
\def\l@subsubsection#1#2{}
\makeatother

\newcommand{\jhu}{William H.\ Miller III Department of Physics and Astronomy, Johns Hopkins University, \\ 3400 North Charles Street, Baltimore, Maryland, 21218, USA}
\newcommand{\UPenn}{Center for Particle Cosmology, Department of Physics and Astronomy, University of Pennsylvania, 209 South 33rd Street, Philadelphia, Pennsylvania 19104, USA}

\begin{document}

\title{Hidden symmetries for tidal Love numbers: \\ Generalities and applications to analog black holes}

\author{Valerio De Luca}
\email{vdeluca2@jh.edu}
\affiliation{\jhu}

\author{Brandon Khek}
\email{bnkhek@sas.upenn.edu}
\affiliation{\UPenn}

\author{Justin Khoury}
\email{jkhoury@upenn.edu}
\affiliation{\UPenn}

\author{Mark Trodden}
\email{trodden@upenn.edu}
\affiliation{\UPenn}


\begin{abstract}
\noindent
Tidal Love numbers characterize the conservative, static response of compact objects to external tidal fields. Remarkably, these quantities vanish identically for asymptotically flat black holes in four-dimensional general relativity. This behavior has been attributed to hidden symmetries—both geometric and algebraic—governing perturbations in these space-times. Interestingly, a similar vanishing of selected multipolar Love numbers arises in the context of supersonic acoustic flows. These systems share several key features with black holes in general relativity, such as the presence of an effective acoustic horizon and a wave equation describing linear excitations. In this work, we explore a symmetry-based connection between the two frameworks and demonstrate that the ladder symmetries observed in acoustic black holes can be traced to structural properties of the underlying wave equation, mirroring those found in general relativistic black hole space-times. 
\end{abstract}

\maketitle

\section{Introduction}
\label{sec:intro}
\noindent
Tidal Love numbers (TLNs) are a set of coefficients that parametrize the conservative tidal response of self-gravitating bodies to external perturbations~\cite{1909MNRAS..69..476L}. Originally defined in Newtonian gravity, TLNs have been extended to general relativity (GR) and have been shown to leave imprints on the gravitational waveforms of binary systems~\cite{Hinderer:2007mb, Binnington:2009bb, Damour:2009vw}. Future observations from detectors such as LISA, the Einstein Telescope, and Cosmic Explorer~\cite{Punturo:2010zz, Sathyaprakash:2019yqt, Maggiore:2019uih, Reitze:2019iox, Kalogera:2021bya, Branchesi:2023mws, LISA:2017pwj, Colpi:2024xhw}, combined with precise tidal modeling, will shed light on the internal structure of compact objects---ranging from the equations of state of neutron stars~\cite{GuerraChaves:2019foa, Chatziioannou:2020pqz} to the physics of black hole horizons---possibly hinting toward the existence of new physics~\cite{Maselli:2018fay, Datta:2021hvm, Pani:2015tga, Cardoso:2017cfl, Nair:2022xfm, Avitan:2023txy, Coviello:2025pla, Pereniguez:2025jxq, Gounis:2025tmt}, perhaps in the gravitational sector~\cite{Cardoso:2017cfl, Cardoso:2018ptl, DeLuca:2022tkm, Barura:2024uog}. 

One of the most striking results regarding tidal interactions in asymptotically flat black holes  within four-dimensional GR is the exact vanishing of their TLNs~\cite{Deruelle:1984hq,Binnington:2009bb,Damour:2009vw,Damour:2009va,Pani:2015hfa,Pani:2015nua,Gurlebeck:2015xpa,Porto:2016zng,LeTiec:2020spy,Chia:2020yla,LeTiec:2020bos,Hui:2020xxx,Charalambous:2021mea,Charalambous:2021kcz,Creci:2021rkz,Bonelli:2021uvf,Ivanov:2022hlo,Charalambous:2022rre,Katagiri:2022vyz,Ivanov:2022qqt,Berens:2022ebl,Bhatt:2023zsy,Sharma:2024hlz,Rai:2024lho}. This result has been confirmed beyond the linear regime~\cite{DeLuca:2023mio,Riva:2023rcm,Iteanu:2024dvx}, and in non-linear GR for static, axisymmetric space-times~\cite{Kehagias:2024rtz,Combaluzier-Szteinsznaider:2024sgb,Gounis:2024hcm}. Recently, the vanishing and non-renormalization of static TLNs for Schwarzschild black holes was shown at fully non-linear order in GR~\cite{Parra-Martinez:2025bcu}. In higher dimensions, the vanishing persists only for specific angular multipoles~\cite{Chakravarti:2018vlt, Chakravarti:2019aup, Pereniguez:2021xcj, Kol:2011vg,Cardoso:2019vof,Hui:2020xxx,Rodriguez:2023xjd,Charalambous:2023jgq,Charalambous:2024tdj,Charalambous:2024gpf,Ma:2024few,Parra-Martinez:2025bcu, Berens:2025jfs, Rodriguez:2025ala}. The behavior of TLNs has also been explored in time-dependent tidal fields~\cite{Bhatt:2024yyz, Ivanov:2022qqt, Charalambous:2022rre, Saketh:2023bul,Perry:2023wmm,Chakraborty:2023zed,Ivanov:2024sds,DeLuca:2024ufn,Bhatt:2024yyz,Katagiri:2024wbg,Katagiri:2024fpn,Chakraborty:2025wvs,Combaluzier--Szteinsznaider:2025eoc, Caron-Huot:2025tlq,Kobayashi:2025vgl, Kosmopoulos:2025rfj}, and in environments with surrounding matter~\cite{Baumann:2018vus,DeLuca:2021ite,DeLuca:2022xlz,Brito:2023pyl,Capuano:2024qhv,Cardoso:2019upw,Cardoso:2021wlq,Katagiri:2023yzm,DeLuca:2024uju,Cannizzaro:2024fpz, DeLuca:2025bph}.

A considerable body of work has sought to explain the vanishing of black hole TLNs through underlying symmetry principles, which we discuss in greater detail later in this manuscript. These so-called hidden symmetries are not isometries of the background space-time but instead emerge at the level of the equations of motion. An algebraic approach was developed in \cite{Charalambous:2021kcz, Charalambous:2022rre}, where an SL$(2,\mathbb{R})$ structure was uncovered in the regime of slowly varying tidal perturbations. A similar algebraic structure also appears in \cite{BenAchour:2022uqo}, though derived from a different viewpoint---namely, as a manifestation of a hidden coordinate invariance that gives rise to ladder operators. Nonlinear extensions of these symmetries have been identified in~\cite{Combaluzier-Szteinsznaider:2024sgb, Kehagias:2024rtz,Gounis:2024hcm,Parra-Martinez:2025bcu}, where they take the form of a generalized Geroch symmetry~\cite{Geroch:1970nt, Breitenlohner:1986um, Maison:2000fj, Lu:2007zv, Lu:2007jc, Katsimpouri:2015nqc}. An alternative line of work has focused on reproducing the same equations of motion by embedding the problem in an effective space-time geometry, such as~$\mathrm{AdS_2 \times S^2}$ \cite{Hui:2022vbh} or a conformally flat background \cite{Hui:2021vcv, Berens:2022ebl}. From this geometrical perspective, the appearance of conformal Killing vectors in these auxiliary space-times---particularly~$\mathrm{AdS_2}$---has been analyzed as a possible origin of the observed structure, as in \cite{Katagiri:2022vyz}.

Despite significant theoretical and observational advances, a complete understanding of black hole physics, even at the classical level, remains incomplete. This has motivated the development of analog gravity models, particularly acoustic black holes (ABHs); see~\cite{Visser:1997ux,Barcelo:2005fc} for reviews. As first shown by Unruh~\cite{Unruh:1980cg}, certain features of black hole space-times can be mimicked by supersonic acoustic flows. In a fluid moving through a tube with varying cross section, the velocity can exceed the local speed of sound, giving rise to an acoustic horizon---an analog of the event horizon in GR. ABHs reproduce many black hole phenomena, including quasinormal ringing, superradiance, tail decay, and even Hawking radiation~\cite{Fischer:2001jz, Basak:2002aw,Berti:2004ju,Cardoso:2004fi,Cardoso:2005ij,Lepe:2004kv,Kim:2004sf,Saavedra:2005ug,Abdalla:2007dz,Vieira:2014rva,Vieira:2021xqw,Vieira:2021ozg,Baak:2023zjf, Singh:2024qfw}. Recently,  we computed in Ref.~\cite{DeLuca:2024nih} the TLNs of ABHs in 2+1 and
in 3+1 dimensions, by studying the excitation of acoustic disturbances in the fluid flow induced by an external
tidal field. We showed that these reproduce a number of properties of higher-dimensional black holes: logarithmic running with radial distance, vanishing for certain angular multipole moments, and the existence of a ladder structure in the perturbation equations, thereby exhibiting a strong similarity with GR black holes~\cite{Hui:2021vcv,Berens:2022ebl,Rai:2024lho, Combaluzier-Szteinsznaider:2024sgb}. 

The main purpose of this work is to continue the investigation started in Ref.~\cite{DeLuca:2024nih} and to deepen the understanding of the relation between black holes and ABHs based on their tidal response. The outline of the paper is as follows. In Sec.~\ref{sec: Schwarzschild} we review the general understanding behind the hidden symmetries for the TLNs of Schwarzschild black holes.  In Sec.~\ref{sec: ABH} we review the main results of Ref.~\cite{DeLuca:2024nih} concerning the tidal deformability in models of analog gravity, identifying the existence of a ladder symmetry of the perturbation equations, which underlies the vanishing of TLNs at special multipole values. In Sec.~\ref{sec: CCKV-ABH} we provide a geometric interpretation of the ladder operators in terms of closed conformal Killing vectors of an effective two-dimensional space-time.
In Secs.~\ref{sec: M\"obius} and~\ref{sec: susy} we discuss the existence of ``hidden'' symmetries behind the origin of ladder operators for both black holes in GR and in analog gravity, showing that they arise from either M\"obius or Darboux transformations.
In Sec.~\ref{sec: Near-zone} we generalize the discussion to time-dependent fields in the near-zone approximation, and show the existence of an effective geometry which is asymptotically~${\rm AdS}_2 \times {\rm S}^2$. The conclusions are left to Sec.~\ref{conclusions}. Appendix~\ref{app vorticity} is devoted to the addition of a mass term in the wave equation, through the inclusion of vorticity in the study of perturbations on ABHs, with the purpose of showing the fragile nature of the TLNs vanishing property.  In the following, we use geometrical units~$G = c = 1$, and mostly positive metric signature.

\section{Hidden symmetries for Schwarzschild black holes}
\label{sec: Schwarzschild}
\noindent
In this Section we provide an overview of the hidden symmetries behind the vanishing of TLNs for~$D$-dimensional Schwarzschild black holes. Their metric takes the form
\begin{align}
\label{metric Schw}
    \mathrm{d}s^2 = - \frac{\Delta(r)}{r^2}  \mathrm{d}t^2 + \frac{r^2\mathrm{d}r^2}{\Delta(r)}+r^2 \mathrm{d}\Omega_{D-2}\,, 
    \end{align}
where
\begin{align}
     \Delta(r)=\frac{1}{r^{D-5}}\Big(r^{D-3}-r_+^{D-3}\Big)\,,
\label{metric Schw Delta}
\end{align}
in terms of the black hole horizon~$r_+$, and the line element~$\mathrm{d}\Omega_{D-2}$ of the unit~$D-2$-sphere. 
Consider a massless, minimally coupled, time-independent scalar field~$\phi(\vec{x})$ on this background. Expanding the scalar in terms of spherical harmonics as\footnote{Given the spherical symmetry of the background, the radial wave function only depends on~$\ell$.}
\begin{equation}
\phi(\vec x) = \sum_{\ell, m} R_{\ell} (r) Y_{\ell m}(\theta,\varphi)\,, 
\label{phi expand}
\end{equation}
the radial wave function~$R_{\ell} (r)$ satisfies the equation of motion
\begin{equation}\label{eq:schwarzschild_radial_waveeqn}
H_\ell  R_{\ell}=0\,,
\end{equation}
where~$H_\ell$ is the Hamiltonian operator
\begin{align}
\label{HstaticSch}
    H_\ell = -r^{D-4}\Delta(r) & \Big[\partial_r\big(\Delta(r) r^{D-4} \partial_r \big) \nonumber \\
    & \,\,\, - r^{D-4} \ell(\ell+D-3)\Big]\,.
\end{align}
This equation admits a hypergeometric solution which, after imposing the condition of regularity at the horizon, behaves at large distances as~\cite{Hui:2020xxx}
\begin{align}
\label{Rasympto}
R_{\ell} (r \to \infty) & \simeq  C \left[ \left( \frac{r}{r_+} \right)^{\ell} + \dots \right. \nonumber \\
& \left.~~~~~~~~~~~~ +~ k^\text{\tiny BH}_{\ell m} \left( \frac{r_+}{r} \right)^{\ell+D-3} + \dots \right]\,.
\end{align}
The first term in brackets, which grows as~$r^{\ell}$, is interpreted as the applied scalar field profile with overall amplitude~$C$. The second term, which falls off as~$r^{-\ell - D + 3}$, encodes the black hole response. 

Following the Newtonian definition of static TLNs~$k^\text{\tiny BH}_{\ell m}$ as the ratio of the response term over the external source~\cite{Deruelle:1984hq,Binnington:2009bb, Damour:2009vw}, these coefficients can be read off from Eq.~\eqref{Rasympto} as taking the values~\cite{Hui:2020xxx} 
\begin{equation}
\label{kBH D} k^\text{\tiny BH}_{\ell} =
\begin{cases}
 \frac{2\hat{\ell}+1}{2\pi} \frac{\Gamma (\hat{\ell}+1)^4}{\Gamma (2\hat{\ell} + 2)^2} \tan (\pi \hat{\ell}) \qquad \quad \,\,\,\,\, \text{for generic} \,\, \hat{\ell}  \\[7pt]
 \frac{(-1)^{2\hat{\ell}} (D-3) \Gamma (\hat{\ell}+1)^2}{(2\hat{\ell})! (2\hat{\ell}+1)! \Gamma(-\hat{\ell})^2} \log \left( \frac{r_0}{r} \right) \,\,\,\,\,\,\,\,\,\, \text{for half-integer} \,\, \hat{\ell}  \\[7pt]
 0 \qquad \qquad \qquad  \qquad \qquad \qquad \,\,\,\, \text{for integer} \,\, \hat{\ell} \,,
\end{cases}
\end{equation}
where
\be
\hat{\ell} \equiv \frac{\ell}{D-3} 
\ee
is an effective multipole moment. In four dimensions~($D = 4$),~$\hat{\ell}= \ell$ takes integer values, and one recovers the well-known result that black hole TLNs vanish for every integer value of~$\ell$. In higher dimensions, however, the static TLNs vanish only when~$\hat{\ell}$ is an integer multiple of~$D - 3$. Interestingly, for half-integer values of~$\hat{\ell}$, the tidal response develops a logarithmic term, which is an example of classical renormalization group running with respect to an arbitrary reference scale~$r_0$~\cite{Kol:2011vg}.

The vanishing of TLNs for certain multipole moments can be established in any space-time dimension by introducing ladder operators~\cite{Hui:2021vcv, BenAchour:2022uqo, Berens:2022ebl, Katagiri:2022vyz}. The static Hamiltonian of Eq.~\eqref{HstaticSch} can be built out of the following ladder operators~\cite{Hui:2021vcv, Berens:2022ebl, Berens:2025jfs}
\begin{align}
\label{Ladder Schw}
        D^+_\ell&= -r^{D-4} \Delta(r) \partial_r + (\ell+D-3)\left(\frac{1}{2}r_+^{D-3} -  r^{D-3}\right)\,; \nonumber \\
    D^-_\ell &= r^{D-4} \Delta(r) \partial_r + \ell\left(\frac{1}{2}r_+^{D-3} - r^{D-3}\right)\,,
\end{align}
satisfying the commutation relations
\begin{align}
\label{HD-BH}
    H_{\ell \pm (D-3)} D^\pm_\ell = D^\pm_\ell H_\ell\,.
\end{align}
It follows that, given a solution~$R_{\ell}$ at level~$\ell$, one can generate solutions at levels~$\ell \pm (D-3)$ by applying the raising/lowering operator. Taking this ladder structure, we can then identify the ground state solution with vanishing TLNs, such that every level obtained by acting with raising/lowering operators will have the same property. Specifically, for~$D$-dimensional non-spinning black holes, the ground state is the level~$\ell=0$, which admits a constant regular solution,~$R_{0} = {\rm constant}$, so that the property of vanishing TLN for this level can be extended to all solutions with multipole values~$\ell \pm n(D-3)$, with~$n \in \mathbb{N}$.

Schwarzschild black holes in~$D$ dimensions also present a  ``horizontal'' symmetry structure, in addition to the ``vertical'' ladder highlighted above~\cite{Berens:2022ebl, BenAchour:2022uqo}. Consider the~$\ell=0$ level operator
\begin{equation}
Q_0\equiv D_0^-=r^{D-4}\Delta\partial_r\,, 
\end{equation}
which commutes with the~$\ell=0$ Hamiltonian,~$[Q_0,H_0] = 0$.  
Such conservation gives rise to a  current $P_0 \equiv Q_0 R_0$ which, upon using the~$\ell=0$ equation of motion, clearly satisfies $\partial_r P_0 = \partial_r(r^{D-4}\Delta \partial_r R_0)=0$, and is thus conserved. 
One can then climb the ladder by constructing the operators~\cite{Berens:2022ebl, BenAchour:2022uqo}
\begin{align}
    Q_\ell= D_{\ell-D+3}^+Q_{\ell-D+3} D_\ell^-\,, \quad \text{with}\quad \left[Q_\ell, H_\ell \right]=0\,,
\end{align}
where $\ell$ is a multiple of $D-3$.  
The conserved quantity~$P_0$ can then be similarly generalized to any~$\ell$ as
\begin{equation}
P_\ell \equiv r^{D-4}\Delta\partial_r\big(D_{D-3}^-\dots D_\ell^-R_\ell\big)\,,
\end{equation}
which is conserved along the radial direction, $\partial_r P_\ell =0$, following the argument discussed above.

Let us now consider a mode function~$R_\ell$ which is regular at the horizon $r_+$. Then, the sequence of lowering operators above maps $R_\ell$ to another regular solution which, because of the dependence of $\Delta(r)$,  enforces  $P_\ell$ to vanish at the horizon,
\be
P_\ell|_{r=r_+}=0\,.
\ee
Since~$P_0$ (and actually any $P_\ell$) is conserved along the radial direction, it follows that~$R_0$ is a constant solution radially, including at spatial infinity. Because
\be
R_\ell=D_{\ell-D+3}^+\dots D_{D-3}^+ D_0^+R_0\,,
\ee
is built from a constant, we can thus infer from the explicit form of~$D_\ell^+$ that~$R_\ell$ does not have decaying terms, enforcing the condition of zero TLNs~\cite{Berens:2022ebl, BenAchour:2022uqo, Berens:2025jfs}. A more detailed discussion of the horizontal ladder and the existence of M\"obius symmetry transformations is left to Sec.~\ref{sec: M\"obius}.

\begin{figure*}[t!]
	\centering
        \includegraphics[width=0.99\textwidth]{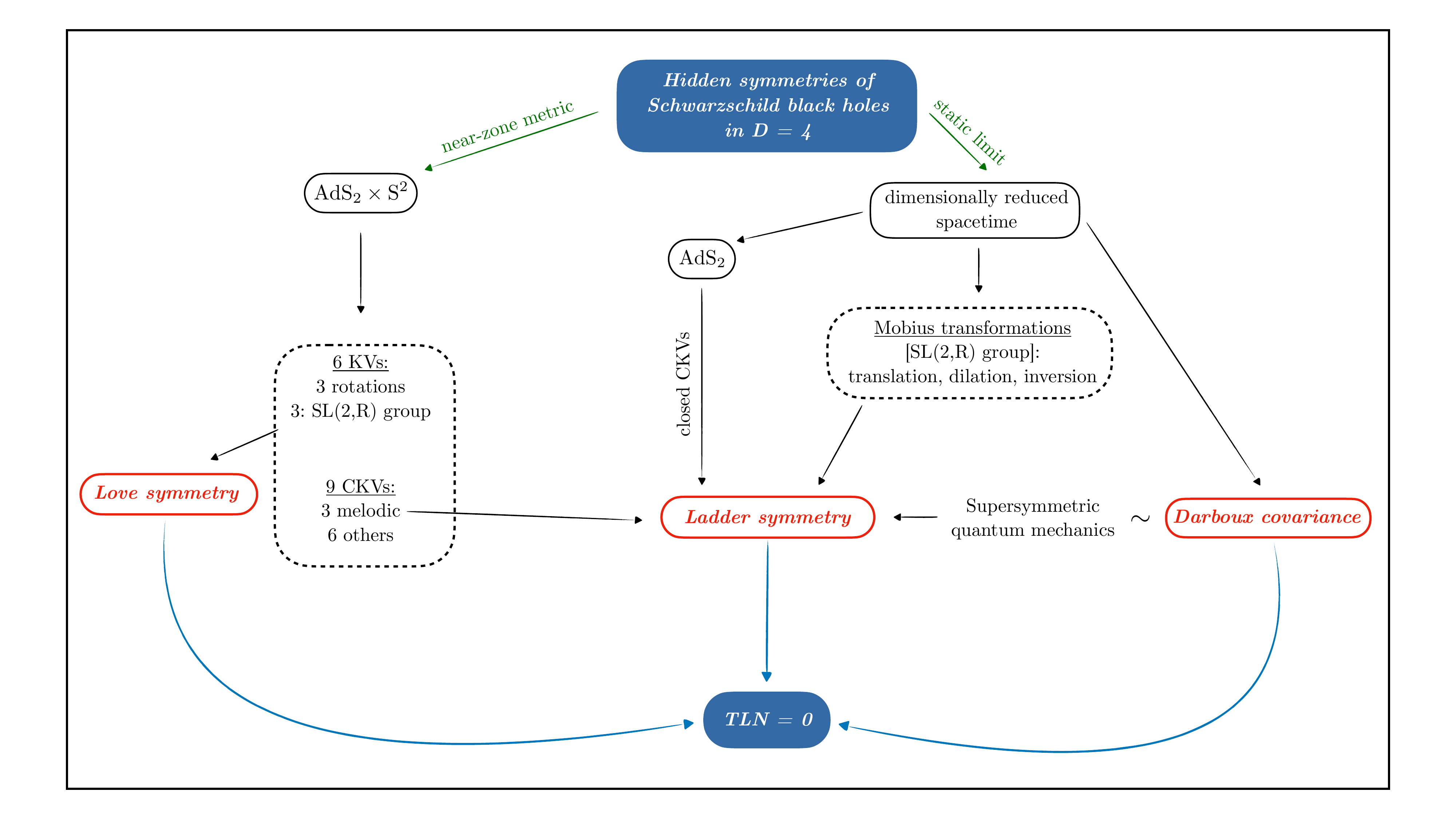}
	\caption{\it Schematic map of the relations among the hidden symmetries behind the vanishing of the TLNs for Schwarzschild black holes in $D = 4$ dimensions. In this paper, we show that a similar diagram also exists for ABHs in the static limit (except that closed CKVs are associated to a 1+1 effective geometry rather than ${\rm AdS}_2$).}
	\label{fig: symmSch}
\end{figure*}

In the following we summarize various results explaining the geometrical and algebraic origin of these ladder structures, which have been explored for~$D$-dimensional black holes and more specifically for the~$D=4$ case. The network of relations among these results for~$D=4$ is schematically shown in Fig.~\ref{fig: symmSch}.

\subsection{Conformal Killing vectors}
\label{sec: CCKV}
\noindent
Let us start by reviewing a geometric interpretation of the ladder operators. In Refs.~\cite{Cardoso:2017qmj, Cardoso:2017egd}, it was argued that there exist ``mass'' ladder operators that map solutions of the massive Klein-Gordon equation of a scalar field~$\phi$ on a certain background to solutions with a different mass. 

In general, suppose~$\phi$ satisfies the equation 
\begin{equation}
\label{2d EOM}
\big(\square -m^2\big)\phi=0\,,
\end{equation}
where~$\Box$ is the d'Alembertian operator of the background metric. Introducing a suitable mass ladder operator~$\mathcal{D}$, one finds that~$\mathcal{D}\phi$ satisfies the mass-shifted equation:
\begin{align}
\label{mass shift}
    \Big(\square - \big(m^2 + \delta m^2\big)\Big)\mathcal{D}\phi =0\,.
\end{align}
In what follows, we explore two examples that use this result to understand the vanishing of the TLNs of Schwarzschild black holes. 
An extended derivation of these results, together with its application to ABHs, is left to Sec.~\ref{sec: CCKV-ABH}.

\subsubsection{Closed conformal Killing vectors of~$\mathrm{AdS_2}$}
\label{ckv of ads_2}
\noindent
It was recognized in~\cite{Katagiri:2022vyz} that the radial equation~\eqref{HstaticSch} of a static scalar field on
a~$D$-dimensional Schwarzschild background coincides with the scalar equation on the~$\mathrm{AdS}_2$ metric
\begin{align}
\label{metricCCKV}
    {\rm d}s_{\mathrm{AdS_2}}^2 = - \Delta(z){\rm d}t^2 + \frac{1}{\Delta(z)}{\rm d}z^2\,,
\end{align}
with~$z=r^{D-3}/r_+^{D-3}$ and~$\Delta(z)=z(z-1)$. In other words, Eq.~\eqref{HstaticSch} can be written as
\begin{align}
    \left[\square_{\mathrm{AdS_2}} - \hat \ell (\hat \ell +1)\right] R_{\ell} = 0\,,
\end{align}
where~$\square_{\mathrm{AdS_2}}=-\frac{1}{\Delta}\partial_t^2 +\partial_z\big(\Delta(z) \partial_z\big)$ and, as before,~$\hat \ell = \frac{\ell}{D-3}$. The~$\mathrm{AdS_2}$ space-time admits three {\it closed conformal Killing vectors} (closed CKVs), vector fields~$\xi^a$ satisfying both the conformal Killing equation
\begin{equation}
\label{CCKV def}
\nabla_a \xi_b + \nabla_b \xi_a = (\nabla_c \xi^c) g_{ab}\,,
\end{equation}
and the closure condition
\be
\nabla_a \xi_b  = \nabla_b \xi_a\,.
\label{closure}
\ee 
Among these, one is relevant in explaining the ladder structure, namely
\begin{align}
    \xi^a \partial_a = \Delta(z) \frac{\partial}{\partial z}\, .
\end{align}
Since~$\xi_a = \partial_a z$, this manifestly satisfies~\eqref{closure}. This closed CKV generates spatial translations in the  coordinate~${\rm d}w = {\rm d}z/\Delta(z)$.
This allows us to define the mass ladder operators
\begin{equation}
\mathcal{D}^\pm_{\ell} = \Delta(z) \frac{\partial}{\partial z} - \frac{k_\pm}{2} (2z-1)\,,
\end{equation}
in terms of~$k_+ = -\hat\ell -1$ and~$k_- = \hat\ell$. It satisfies the commutation relation
\begin{align}
&\big(\mathcal{D}^\pm_\ell + \nabla_a \xi^a\big) \left[\square_{\mathrm{AdS_2}} - k_\pm(k_\pm+1) \right] R_{\ell} \nonumber \\
& ~~~~~~~~~~~~~~~~~~~~~~~=\left[\square_{\mathrm{AdS_2}} - k_\pm (k_\pm- 1)\right]\mathcal{D}^\pm_\ell R_{\ell}\,.
\end{align}
In other words, the mass ladder operator associated to the closed CKV corresponds to the ladder operator~$D_\ell^\pm$ of Eq.~\eqref{Ladder Schw}, where the ``mass'' of the radial equation is merely the~$\ell$ mode. Ref.~\cite{Katagiri:2022vyz} also pointed out a supersymmetric structure in the~$D$-dimensional case, besides the hidden space-time conformal symmetry. We will explore this idea later in Sec.~\ref{sec: susy}.

\subsubsection{Melodic conformal Killing vectors}
\noindent
The case with vanishing mass term,~$m^2=\delta m^2=0$, corresponding to a massless Klein-Gordon equation,~$\square \phi =0$,
was also considered in~\cite{Cardoso:2017egd}. In this case, the mass ladder operators are usually identified as ``symmetry operators''. Indeed, it was shown that, if there exists a (not necessarily closed) CKV~$\xi^\mu$ that satisfies the condition
\begin{align}
\label{eq:melodic}
    \Box \nabla_\mu \xi^\mu=0\,,
\end{align}
then the ladder operator 
\begin{align}
    \mathcal{D}=\xi^\mu \nabla_\mu + \frac{D-2}{2D}\nabla_\mu \xi^\mu
\end{align}
is a symmetry operator of the Klein-Gordon equation. 
These results correspond  exactly to those obtained in Ref.~\cite{Berens:2022ebl}, where the free scalar field action, acted upon by a CKV $\xi^\mu$ of the effective $D-1$ dimensional metric
\begin{align}
\label{metriceff3}
    \d s^2 = \d r^2 + \Delta(r)\d \Omega_2\,,
\end{align}
is invariant only when the ``melodic condition'' of Eq.~\eqref{eq:melodic} is satisfied, 
where the covariant derivatives shown are compatible with this metric. The effective metric~\eqref{metriceff3} reproduces the static radial wave equation in Eq.~\eqref{eq:schwarzschild_radial_waveeqn}. The variation of the scalar field under $\xi^\mu$ is obtained by acting with the operator~$\mathcal{D}$ shown above, recovering the known ladder operators~$D_\ell^\pm$~\cite{Berens:2022ebl}. A more focused discussion of melodic CKVs has recently been presented in Ref.~\cite{Berens:2025okm}.

\subsection{Near-zone symmetries}
\label{sec: near zone symmetries}
\noindent
While the above discussion has focused on time-independent perturbations, it has been shown in Ref.~\cite{Hui:2022vbh} that, when considering long-wavelength perturbations on Schwarzschild black holes, a larger group of symmetry transformations appears compared to the strictly static limit. In particular, in frequency space~$(\phi \propto {\rm e}^{-{\rm i} \omega t})$, the behavior of~$\phi$ in the near-zone region, defined by~$r_+ \leq r \ll 1/\omega$, is described through the effective metric (in~$D= 4$ dimensions)
\begin{equation}
\label{NZ metric}
{\rm d}s^2_\text{\tiny nz} = - \frac{\Delta(r)}{r_+^2}  \mathrm{d}t^2 + \frac{r_+^2\mathrm{d}r^2}{\Delta(r)}+r_+^2 \mathrm{d}\Omega_{2}\,.
\end{equation}
The scalar equation of motion on this effective metric coincides with that on the original Schwarzschild space-time in the static limit ($\omega = 0$). The effective metric~\eqref{NZ metric} is recognized as AdS$_2 \times S^2$. Since this geometry has vanishing Ricci scalar, the equation for~$\phi$ coincides with that of a conformally coupled scalar.

The~AdS$_2 \times S^2$ space-time has six KVs (compared to the usual four of Schwarzschild), which include the usual three rotations plus
\begin{align}
\label{SL2RLove}
T  = 2 r_+ \partial_t \,, \quad   
L_{\pm} = e^{\pm \frac{t}{2r_+}} (2r_+ \partial_r \sqrt{\Delta} \partial_t \mp \sqrt{\Delta} \partial_r)\,.
\end{align}
These three KVs, introduced in~\cite{Charalambous:2021kcz,Charalambous:2022rre,Charalambous:2024gpf}, form a~$\mathfrak{sl}(2,\mathbb{R})$ subalgebra 
\begin{equation}
[T, L_\pm] = \pm L_\pm\,; \qquad [L_+,L_-] = -2 T\,.
\end{equation}
The Casimir operator of this algebra,
\be
H = -T^2 + \frac{1}{2} \big(L_+ L_- + L_- L_+\big)\,,
\ee
coincides with the wave operator of the scalar field on the effective metric~\eqref{NZ metric}. This~SL$(2,\mathbb{R})$ group has been identified in the literature as the {\it Love symmetry}, providing an explanation of the vanishing of the static TLNs. In particular, static perturbations belong to highest weight representations of~$\mathfrak{sl}(2,\mathbb{R})$, which forces the static TLNs to vanish~\cite{Charalambous:2021kcz,Charalambous:2022rre,Charalambous:2024gpf}.

The near-zone metric~\eqref{NZ metric}, being conformally flat, also contains nine CKVs, which include
\begin{align}
J_{01} & = - \frac{2 \Delta}{r_+} \cos \theta \, \partial_r - \frac{\partial_r \Delta}{r_+} \sin \theta \, \partial_\theta\,, \nonumber \\
J_{02} & = - \cos \varphi \, \left[ \frac{2 \Delta}{r_+} \sin \theta \, \partial_r + \frac{\partial_r \Delta}{r_+} \left(\frac{\tan \varphi}{\sin \theta} \partial_\varphi -  \cos \theta \, \partial_\theta \right) \right]\,, \nonumber \\
J_{03} & = - \sin \varphi \, \left[ \frac{2 \Delta}{r_+} \sin \theta \, \partial_r - \frac{\partial_r \Delta}{r_+} \left(\frac{\cot \varphi}{\sin \theta} \partial_\varphi + \cos \theta \, \partial_\theta \right) \right]\,.
\end{align}
These three CKVs generate symmetries of the exact system in the static limit and coincide with the melodic CKVs discussed in Ref.~\cite{Berens:2022ebl}. In particular, the first operator, when acting on a static solution, generates the ladder operators $D^\pm_\ell$ discussed in Eq.~\eqref{Ladder Schw}. This gives a geometric interpretation of the ladder operator in the near-zone effective space-time. 

\section{Ladder structure of analog black holes}
\label{sec: ABH}
\noindent
In this Section, we briefly summarize the results for the~$3+1$-dimensional canonical ABH found in Ref.~\cite{DeLuca:2024nih}. 
We consider the most general spherically symmetric acoustic metric, associated to the flow of an incompressible fluid with constant pressure and sound speed:
 \begin{align}\label{eq:canonical_metric}
    \mathrm{d}s^2 = - \frac{\Delta(r)}{r^2} c_s^2  \mathrm{d}t^2 + \frac{r^2 \mathrm{d}r^2}{\Delta(r)}+r^2 \mathrm{d}\Omega_2\,,
\end{align}
with 
\begin{align}\label{eq:canonical_fr}
\Delta(r) = r^2 \left(1-\frac{r_+^4}{r^4}\right)\,.
\end{align}
By studying a massless test scalar propagating on this metric, one can derive the corresponding static TLNs. These were found in~\cite{DeLuca:2024nih} to vanish for multipoles~$\ell = 4n$ and~$3 + 4n$, where~$n$ is an integer. As we review below, this can be understood in terms of a ladder structure, in complete analogy with that for~$3+1$-dimensional Schwarzschild black holes in GR.

\subsection{Acoustic TLNs}
\noindent
Consider a massless scalar field~$\phi$ on the acoustic metric~\eqref{eq:canonical_metric}. Expanding the field in terms of spherical harmonics, as in Eq.~\eqref{phi expand}, 
the radial wave function~$R_{\ell} (r)$ satisfies 
\begin{align}
\label{eq:canonical_radial_eom}
    \Big(\Delta(r)R_{\ell}'(r)\Big)'-\ell(\ell+1) R_{\ell}(r)=0\,.
\end{align}
As shown in~\cite{DeLuca:2024nih}, the general solution is in terms of hypergeometric functions. The solution that is regular at the ABH horizon~$r = r_+$ is
\begin{equation}
\label{Rcan}
R_{\ell} = C \left(\frac{r}{r_+}\right)^\ell {}_2 F_1\left(a,b,c;\, 1-\frac{r_+^4}{r^4}\right)\,,
\end{equation}
where~$C$ is a constant, and 
\begin{equation}
    a=\frac{3-\ell}{4}\,; \qquad b=-\frac{\ell}{4}\,; \qquad c = 1\,.
\end{equation}
By studying its asymptotic behavior at large distances, one can identify the growing and decaying modes, in analogy with Eq.~\eqref{Rasympto}, and extract the TLNs as~\cite{DeLuca:2024nih}
\begin{align}
\label{eq:TLNABH}
    k^\text{\tiny (3+1)}_{\ell} = \frac{\Gamma \left(-\frac{\ell}{2}-\frac{1}{4}\right) \Gamma \left(\frac{\ell}{4}+1\right) \Gamma \left(\frac{\ell+1}{4}\right)}{\Gamma \left(\frac{3}{4}-\frac{\ell}{4}\right) \Gamma \left(\frac{\ell}{2}+\frac{1}{4}\right) \Gamma \left(-\frac{\ell}{4}\right)}\,.
\end{align}
Notably, this vanishes for~$\ell=4n$ and~$3 + 4n$, where~$n\in \mathbb{N}$, similarly to~seven-dimensional Schwarzschild black holes. As discussed in~\cite{DeLuca:2024nih}, this property can be traced back to the fact that both metrics have the same radial dependence in the metric function~$\Delta (r)$. [Compare~\eqref{eq:canonical_fr} and Eq.~\eqref{metric Schw Delta} with~$D = 7$.]
However, contrary to Schwarzschild black holes, for which there is a unique ground state level~$\ell = 0$ with vanishing TLN, here the two states~$\ell = 0$ and~$\ell = 3$ are both ground states over which a ladder structure can be built.  

\subsection{Ladder symmetries}
\noindent
The vanishing of the TLNs for these particular multipoles can be explained by the existence of ladder operators, which relate solutions of the equation of motion at different~$\ell$. Namely, the radial equation~\eqref{eq:canonical_radial_eom} can be written as
\be
H_\ell R_{\ell}=0\,, 
\label{EOM-ABH 0}
\ee
with
\begin{equation}
\label{EOM-ABH}
    H_\ell \equiv - r^6 \left[\Delta (r) \frac{{\rm d}}{{\rm d}r} \left(\Delta (r) \frac{{\rm d}}{{\rm d}r}  \right)- \ell (\ell+1) \Delta (r) \right]\,.
\end{equation}
Introducing the ladder operators~\cite{DeLuca:2024nih}
\begin{align}
\label{ladderABH}
D^+_\ell &\equiv -r^3 \Delta (r) \frac{{\rm d}}{{\rm d}r}   - (\ell+1) r^4 +\frac{(\ell+1)^2}{2 \ell+5} r_+^4 \,;  \nonumber \\
D^-_\ell &\equiv r^3 \Delta (r) \frac{{\rm d}}{{\rm d}r}   - \ell r^4 + \frac{\ell^2}{2 \ell-3} r_+^4\,,
\end{align}
one can easily check that
\begin{align}
\label{HDD}
H_\ell & = D^-_{\ell + 4} D^+_\ell - \frac{(\ell+1)^2 (\ell+4)^2}{(2 \ell+5)^2} r_+^8 \nonumber \\
 & =  D^+_{\ell-4} D^-_\ell - \frac{\ell^2  (\ell-3)^2 }{(2 \ell-3)^2} r_+^8 \,,
\end{align}
and
\begin{align}
\label{HD-ABH}
    H_{\ell \pm 4} D^\pm_\ell = D^\pm_\ell H_\ell\,.
\end{align}
Thus, if~$R_{\ell}$ is a solution to~$H_\ell R_{\ell}=0$, then~$D_\ell^\pm R_{\ell}$ satisfies~$H_{\ell\pm 4}D_\ell^\pm  R_{\ell}=0$. In other words, starting from a level-$\ell$ solution~$R_{\ell}$, we can raise it to a level~$\ell+4$ solution through~$D^+_\ell R_{\ell}$, or lower it to a level~$\ell-4$ solution through~$D^-_\ell R_{\ell}$. It follows that~$D^\pm_\ell R_{\ell} = \mathcal{C}_{\ell}^\pm R_{\ell \pm 4}$, with overall constants~$\mathcal{C}_{\ell}^\pm$ which enforce relations between solutions with different~$\ell$.

The discussion so far largely parallels the results obtained for black holes in GR. However, ABHs have an additional structure, in the form of a {\it small ladder}, that allows us to connect the~$\ell = 0$ and~$\ell = 3$ ground states of the theory. This is analogous to the response coefficients of perturbations in theories characterized by screening mechanisms, such as Born-Infeld electromagnetism or Dirac-Born-Infeld scalar theory~\cite{BeltranJimenez:2022hvs, BeltranJimenez:2024zmd}. 

Specifically, one can define the further set of operators
    \begin{align}
    \tilde{D}^+_\ell &\equiv -r^3 \Delta (r) \frac{{\rm d}}{{\rm d}r} + \ell r^4 +\frac{\ell^2}{3-2 \ell} r_+^4 \,;  \nonumber \\
    \tilde{D}^-_\ell &\equiv r^3 \Delta (r) \frac{{\rm d}}{{\rm d}r}  + (\ell-3) r^4 + \frac{(\ell-3)^2}{3-2 \ell} r_+^4\,,
\end{align}
which satisfy the commutation relations
\begin{align}
    H_{3- \ell} \tilde{D}^+_\ell &= \tilde{D}^+_\ell H_\ell \,, \nonumber \\
H_{\ell} \tilde{D}^-_\ell &= \tilde{D}^-_\ell H_{3-\ell}\,.
\end{align}
These allow us to relate only the solutions for the modes~$\ell=0,3$. In this sense, one can draw two parallel big ladders, starting from the two ground states and climbing both branches in steps of~$\Delta \ell=4$~\cite{DeLuca:2024nih}, and a small ladder relating solutions between the two ground states. 
Furthermore, it is interesting that the two set of ladder operators can be related via~$\tilde{D}^+_\ell = - D^-_\ell$ and~$\tilde{D}^-_{\ell+4}  = -D^+_\ell$,
hinting toward a common origin behind these operators, despite satisfying different commutation relations with the Hamiltonian. 

In the next Sections, we provide explanations for the existence of these operators as arising from ``hidden symmetries'', {\it i.e.}, symmetries not manifest at the level of the original action, but uncovered in effective space-times or in the equations of motion in a suitable coordinate system. The upshot is to provide both a geometric and algebraic symmetry understanding of these operators.

\section{Ladder symmetry from closed conformal Killing vectors}
\label{sec: CCKV-ABH}
\noindent
In this Section we provide a geometric understanding of ladder operators for ABHs through the existence of closed CKVs associated to an effective 1+1 dimensional space-time, in analogy to Sec.~\ref{sec: CCKV} for Schwarzschild black holes~\cite{Katagiri:2022vyz}. Even though these results will be applied to ABHs, let us stress that most of the derivation applies straightforwardly to any static, spherically symmetric metric, thus providing a more general result than black holes in GR or analog gravity.

The starting point is to notice that the static radial equation of motion of the free massless scalar on the canonical ABH background, given by Eqs.~\eqref{EOM-ABH 0} and~\eqref{EOM-ABH}, coincides with that of a {\it massive} scalar field in 1+1 dimensions, 
\be
\big(\square - m_\ell^2\big)R_\ell=0\,,
\label{2d EOM bis}
\ee
where~$m_{\ell}^2 = \ell(\ell+1)$, and~$\square$ is the d'Alembertian for the effective geometry (in the following, we set $c_s = 1$ for simplicity)
\begin{align}
\label{metricCCKV bis}
    {\rm d}s_2^2 = - \Delta(r){\rm d}t^2 + \frac{1}{\Delta(r)}{\rm d}r^2\,,
\end{align}
with~$\Delta (r) = r^2 (1- r_+^4/r^4)$. The spatial coordinate was denoted in general by~$z$ in Eq.~\eqref{metricCCKV} but coincides with~$r$ in this case.

Similarly to the results for Schwarzschild black holes, it is possible to determine the closed CKVs of this effective geometry. That is, we seek two-dimensional vector fields~$\xi^a = (\xi^t,\xi^r)$ satisfying the closed conformal Killing equation~\cite{Cardoso:2017qmj, Cardoso:2017egd}\footnote{This follows from combining the conformal Killing equation~\eqref{CCKV def} and closure condition~\eqref{closure}.} 
\begin{equation}
\nabla_a \xi_b = \frac{1}{2} \nabla_c \xi^c g_{ab} \,.
\label{ckv eqn}  
\end{equation}
The components of this equation are explicitly given by
\begin{subequations}
\begin{align}
 \dot{\xi}_t &=  -\Delta^2\xi_r'\,; \\
 \xi_t' &=   \frac{\Delta'}{2 \Delta}\xi_t \,; \\
 \dot{\xi}_r &= \xi_t' \,,
\end{align}
\end{subequations}
where prime denotes differentiation with respect to~$r$. It is easy to show that the unique solution to these equations is
\be
\xi_t = 0\,;\quad \xi_r = 1\quad  \Longrightarrow \quad \xi^t = 0\,;\quad \xi^r = \Delta(r)\,.
\label{cckv}
\ee
The physical interpretation of this diffeomorphism is most transparent in the coordinate~${\rm d} x = \frac{{\rm d}r}{\Delta(r)}$, where the metric~\eqref{metricCCKV bis} is conformally flat 
\be\label{metric x}
{\rm d}s^2_2 = \Delta(r(x)) \left( - {\rm d}t^2 + {\rm d}x^2\right)\,.
\ee
The diffeomorphism of Eq.~\eqref{cckv} is thus just a {\it spatial translation},~$x \rightarrow x + \epsilon$, under which the metric remains conformally flat.

We now wish to generalize the construction of Refs.~\cite{Cardoso:2017qmj, Cardoso:2017egd, Katagiri:2022vyz} to relate the closed CKV identified above to our ladder operators. That is, we look for a set of operators~$\mathcal{D}_\ell$ satisfying
\be
[\Box,\mathcal{D}_\ell] = \delta m_{\ell}^2 \mathcal{D}_\ell + 2Q \big(\Box  - m_{\ell}^2 \big) \,,
\label{D cond}
\ee
where~$\delta m_{\ell}^2$ is a constant to be determined, and~$Q$ is a function of $r$. One could assume $\ell$ dependence, though $Q$ will turn out to be independent of~$\ell$ from the constraints to follow. In this way, if~$R_\ell$ satisfies~\eqref{2d EOM bis}, then~$\mathcal{D}_\ell R_\ell$ will satisfy Eq.~\eqref{mass shift} (reproduced here for convenience):
\begin{align}
\label{mass shift bis}
    \Big(\square - \big(m_{\ell}^2 + \delta m_{\ell}^2\big)\Big)\mathcal{D}_\ell R_\ell =0\,.
\end{align}
In other words,~$\mathcal{D}_\ell$ acts as a ladder operator, with~$\delta m_{\ell}^2$ setting the mass shift. 

Following~\cite{Katagiri:2022vyz}, we make the ansatz
\be
\mathcal{D}_\ell = G \xi^a\nabla_a + K_\ell \,,
\label{D ansatz}
\ee
where~$G$ and~$K_\ell$ are functions of~$r$, and~$\xi^a$ is a closed vector,~$\nabla_a \xi_b = \nabla_b \xi_a$. Since our metric depends only on~$r$, it suffices to assume that~$G$ and~$K_\ell$ are functions of~$r$ only. 
We will find that only~$K$ ends up depending on~$\ell$, which we have denoted above. We will show that imposing Eq.~\eqref{D cond} enforces~$\xi^a$ to be the closed CKV of~\eqref{cckv}. Notice that our ansatz is slightly more general than Ref.~\cite{Katagiri:2022vyz}, which assumes~$G = 1$, as we will see that a non-trivial~$G$ is necessary in our case. In other words, we will provide a more general form for the ladder operator than that assumed in~\cite{Cardoso:2017qmj, Cardoso:2017egd, Katagiri:2022vyz}. 
Demanding that~$\mathcal{D}_\ell$ satisfies Eq.~\eqref{D cond} will allow us to fix the three functions~$G$,~$K_\ell$ and~$Q$, as well as the constant~$\delta m_{\ell}^2$. In the following, d’Alembertian operators simply reduce to Laplacian operators $\nabla^2 \equiv \partial_r(\Delta\partial_r)$, given that we focus on static test fields.  

Acting with~$\mathcal{D}_\ell$ on an arbitrary scalar field~$\psi$, we obtain
\begin{align}
\nonumber
[\nabla^2,\mathcal{D}_\ell]\psi &= \Big(\xi^a \nabla^2 G + 2 \nabla^a K_\ell + G(\nabla^2 \xi^a + R_{~b}^{a} \xi^b)\Big) \nabla_a\psi \nonumber \\
&+ 2 G (\nabla_b \xi^a) \nabla^b\nabla_a \psi + 2(\nabla^b G) \nabla_b (\xi^a  \nabla_a\psi ) \nonumber \\
&   + \nabla^2 K_\ell \psi\,,
\label{com 1}
\end{align} 
where~$R_{~b}^{a}$ is the Ricci tensor of the 2d geometry~\eqref{metricCCKV bis}. It can be shown that Eq.~\eqref{com 1} matches exactly the form of Eq.~\eqref{D cond}, provided that
the following conditions are satisfied:
\begin{subequations}
\begin{align}
\label{conditions prel}
2Q g_{ab} &= 2 G(\nabla_b\xi_a) + 2 \xi_c (\nabla^c G) g_{ab} \,; \\
\label{conditions prel 2}
\nabla_a K_\ell &= \frac{1}{2} \Big(\delta m_{\ell}^2 G - \nabla^2 G\Big)\xi_a - \frac{1}{2}G(\nabla^2 \xi_a + R_{a}^{~b} \xi_b)\,; \\
\nabla^2 K_\ell &= \delta m_{\ell}^2 K - 2m_{\ell}^2 Q\,,
\end{align}
\end{subequations}
where we have simplified the last term in the second line of Eq.~\eqref{com 1} as~$\nabla_b (\xi^a  \nabla_a\psi ) = \xi_b \nabla^2 \psi$ for the metric of Eq.~\eqref{metricCCKV bis}.

The first condition~\eqref{conditions prel} clearly imposes that~$\xi_a$ is a closed CKV, so that~$\xi_a$ satisfies~\eqref{ckv eqn}. In this case every term in~\eqref{conditions prel} is proportional to~$g_{ab}$.
In two dimensions, a closed CKV satisfies~$\Box \xi_a + R_a^{~b} \xi_b = 0$, as can be seen by taking the divergence of~\eqref{ckv eqn} and using the closed condition with the commutator of covariant derivatives. Using these facts, the above conditions reduce to
\begin{subequations}
\begin{align}
\label{conditions}
Q &= \frac{1}{2} G\nabla_c\xi^c + \xi^a\nabla_a G \,; \\
\label{conditions 2}
\nabla_a K_\ell &= \frac{1}{2} \Big(\delta m_{\ell}^2 G - \nabla^2 G\Big)\xi_a\,; \\
\label{conditions 3}
\nabla^2 K_\ell &= \delta m_{\ell}^2 K - 2m_{\ell}^2 Q\,.
\end{align}
\end{subequations}
Taking the divergence of~\eqref{conditions 2}, equating the result to~\eqref{conditions 3}, and eliminating~$Q$ using~\eqref{conditions}, allows us to solve for~$K_\ell$:
\begin{align}
\label{eq: K}
K_\ell &= \frac{1}{\delta m_{\ell}^2} \xi^b\nabla_b \left[ \left(\frac{1}{2} \delta m_{\ell}^2 + 2m_{\ell}^2\right) G - \frac{1}{2} \nabla^2 G\right]  \nonumber \\
& + \frac{1}{\delta m_{\ell}^2} \nabla_c\xi^c \left[\left(\frac{1}{2} \delta m_{\ell}^2 + m_{\ell}^2\right) G - \frac{1}{2} \nabla^2 G\right] \,.
\end{align}
This can then be substituted back into~\eqref{conditions 2} to obtain an equation for~$G$:
\begin{align}
\nonumber
& \nabla_c\xi^c \nabla_a \left[\left(\frac{3}{4} \delta m_{\ell}^2 + 2m_{\ell}^2\right) G - \frac{3}{4} \nabla^2 G\right]  \nonumber \\
& + \nabla_a\big(\nabla_c\xi^c\big) \left[\left(\frac{1}{2} \delta m_{\ell}^2 + m_{\ell}^2\right) G - \frac{1}{2} \nabla^2 G\right] \nonumber \\
& + \xi^b \nabla_b\nabla_a  \left[\left(\frac{1}{2} \delta m_{\ell}^2 + 2m_{\ell}^2\right) G - \frac{1}{2} \nabla^2 G\right]  \nonumber \\
& = \frac{\delta m_{\ell}^2}{2} \Big(\delta m_{\ell}^2G - \nabla^2 G\Big) \xi_a \,.
\end{align}
Explicitly, for the metric~\eqref{metricCCKV bis} and closed CKV~\eqref{cckv}, this gives
\begin{widetext}
\begin{align}
\nonumber
&\left(\delta m_{\ell}^2 + 2m_{\ell}^2\right) \Delta G'' + \left(\frac{3}{2}\delta m_{\ell}^2 + 3m_{\ell}^2 \right) \Delta'G' + \left(\frac{1}{2}\delta m_{\ell}^2 + m_{\ell}^2\right) \Delta''G - \frac{1}{2} \Delta^2G''''  - ~\frac{5}{2} \Delta\Delta'G'''   \nonumber \\
& - 2\left(\Delta'^2 +\Delta\Delta''\right)G'' - \frac{1}{2} \big(3\Delta'\Delta'' + \Delta \Delta'''\big) G' - \frac{(\delta m_{\ell}^2)^2}{2} G  =0 \,.
\label{G ode}
\end{align}
\end{widetext}
This is a fourth-order differential equation for~$G(r)$ which, as stressed at the beginning of this Section, can be applied to any effective metric of the form Eq.~\eqref{metricCCKV bis}.

To solve Eq.~\eqref{G ode}, we make the power-law ansatz
\be
G(z) = r^\alpha\,, 
\ee
with constant~$\alpha$. Together with~$\Delta (r) = r^2 \left(1- \frac{r_+^4}{r^4}\right)$ for canonical ABHs, we obtain 
\begin{widetext}
    \begin{align}
        &\bigg[\Big((\alpha-2   \ell)(\alpha+1) -\delta m^2_{\ell} \Big) \Big((\alpha +1) (\alpha +2 \ell+2)-\delta m^2_{\ell}\Big)        \bigg]r^{8} \nonumber\\
        &+\bigg[(\alpha -3) (\alpha -1)  \Big(\alpha(2-\alpha) +2 \ell(\ell+1)+\delta m^2_{\ell}\Big)\bigg]2r_+^4 r^4 +\bigg[(\alpha -7) (\alpha -6) (\alpha -3) \alpha\bigg] r_+^8 =0\,.
    \end{align}
\end{widetext}
We can solve for $\alpha$ and $\delta m_\ell^2$ by noticing that the equation must vanish term by term, giving us~$\alpha = 3$,~$\delta m_{\ell}^2 = -4 (2\ell -3)$, and~$\delta m_{\ell}^2 = 4 (2 \ell+5)$. These correspond respectively to mass shifts given by
\begin{align}
m_{\ell}^2 &= \ell(\ell+1)  \overset{\delta m_{\ell}^2 = -4 (2\ell -3)}{\Longrightarrow}    m_{\ell}^2 + \delta m_{\ell}^2 = (\ell-4)(\ell-3) \,; \nonumber \\
m_{\ell}^2 &= \ell(\ell+1)  \overset{\delta m_{\ell}^2 = 4 (2\ell +5)}{\Longrightarrow}    m_{\ell}^2 + \delta m_{\ell}^2 = (\ell+4)(\ell+5) \,, 
\end{align}
which therefore shift~$\ell$ to~$\ell - 4$ and $\ell + 4$, respectively.

The solution~$G(r) = r^3$ thus obtained allows us to fix, via Eq.~\eqref{conditions},
\be
Q (z) = 4 r^4 - 2 r_+^4\,.
\ee
Using~$G(r) = r^3$ and the expressions for~$\delta m_{\ell}^2$ in Eq.~\eqref{eq: K} gives the solutions for~$K_\ell$:
\begin{align}
K^-_\ell (z) & = - \ell r^4 + \frac{\ell^2}{2\ell-3}r_+^4\,, \qquad ~~~~\,\delta m_{\ell}^2 = -4 (2\ell -3)\,; \nonumber \\
K^+_\ell (z) & = (\ell+1) r^4 - \frac{(\ell+1)^2}{2\ell+5}r_+^4\,, \quad \delta m_{\ell}^2 =  4 (2 \ell+5)\,.
\end{align}
Lastly, these expressions for~$K_\ell$, together with~$G$ and the closed CKV~$\xi^a$ given by Eq.~\eqref{cckv}, can all be substituted into Eq.~\eqref{D ansatz} to
obtain~${\cal D}_\ell^\pm$:
\begin{align}
{\cal D}_\ell^+ (z) & =  r^3 \xi^a\partial_a + K_\ell^+ \,; \nonumber \\
{\cal D}_\ell^- (z) & = r^3 \xi^a\partial_a + K_\ell^- \,.
\end{align}
We recognize these as the ladder operators for canonical ABHs given in Eq.~\eqref{ladderABH}.
Therefore, the existence of a closed CKV gives a geometric origin for the ladder operators of ABHs, in analogy with the one of Schwarzschild black holes (whose effective metric, we recall, is AdS$_2$).

\section{Ladder symmetry from M\"obius transformations}
\label{sec: M\"obius}
\noindent
In this Section, we investigate the origin of the  ladder symmetry structure from an algebraic perspective and show that it can be obtained from a~$\mathrm{SL}(2,\mathbb{R})$ transformation on a new spatial coordinate~\cite{BenAchour:2022uqo, Hui:2021vcv}. We begin by framing the problem quite generally following Ref.~\cite{BenAchour:2022uqo}, and then specialize to the case of canonical~$3+1$-dimensional ABHs.

Assuming separability of a scalar field with radial eigenfunction~$\psi_\ell(z(r))$, the static wave equation can always be recast into the canonical Sturm-Liouville form ({\it i.e.}, without first-order derivatives)
\begin{align}\label{eq:eomz}
    \left[\frac{{\rm d}^2}{{\rm d} z^2} + \mathcal{V}_\ell(z) \right] \psi_\ell(z) = 0\,,
\end{align}
for some potential~$\mathcal{V}_\ell(z)$. In the case of black holes, the effective potential carries information about the singular points associated to the space-time, such as the black hole horizon and spatial infinity. Equation~\eqref{eq:eomz} admits two independent homogeneous solutions,
\begin{equation}
\label{homo}
\psi_\ell (z) = c_1 \psi_\ell^\mathrm{reg} (z) + c_2 \psi_\ell^\mathrm{irreg} (z)\,,
\end{equation}
where~$ \psi_\ell^\mathrm{reg} (z)$ is regular at the horizon and grows at spatial infinity, while~$\psi_\ell^\mathrm{irreg} (z)$ diverges at the horizon and decays at infinity.

This problem admits Lagrangian and Hamiltonian formulations along the ``time'' direction~$z$, with
\begin{align}
\mathcal{L}_\ell &= \frac{1}{2} \left[\big(\psi'_\ell \big)^2 - \mathcal{V}_\ell \psi^2_\ell \right]\,;  \nonumber \\
\mathcal{H}_\ell &= \Pi_\ell \psi'_\ell - \mathcal{L}_\ell = \frac{1}{2} \Big[\Pi_\ell^2 + \mathcal{V}_\ell \psi_\ell^2 \Big]\,, 
\end{align}
where~$\Pi_\ell = \frac{\delta \mathcal{L}}{\delta \psi'_\ell} = \psi'_\ell$ is the conjugate momentum, and primes now denote derivatives with respect to~$z$.
These quantities satisfy Hamilton's equations
\begin{align}
\psi'_\ell &= \{\psi_\ell,\mathcal{H}_\ell\} = \Pi_\ell,;\\
\Pi'_\ell & = \{\Pi_\ell,\mathcal{H}_\ell\} = - \mathcal{V}_\ell \psi_\ell\,,
\end{align}
where the Poisson bracket is defined by the canonical pair $\{\psi_\ell,\Pi_\ell \} = 1$.

Following Ref.~\cite{BenAchour:2022uqo}, the system is invariant under a Galilean field translation 
\begin{align}
z &\to \tilde{z} = z\,, \nonumber \\
\psi_\ell(z) &\to \tilde{\psi}_\ell (\tilde{z}) = \psi_\ell(z) + \chi_\ell(z)\,,
\end{align}
where~$\chi_\ell(z)$ is the solution to the Sturm-Liouville equation~\eqref{eq:eomz}. Since the theory is quadratic, shifting the field by a solution to the equation of motion is clearly a symmetry, though~$\psi_\ell(z)$ need not be. For this reason, this is an off-shell symmetry. The corresponding Noether charge is~\cite{BenAchour:2022uqo}
\begin{align}\label{eq:noether_chargeY}
Y[\chi_\ell, \psi_\ell] = \delta \psi_\ell \frac{\delta \mathcal{L}_\ell}{\delta \psi'_\ell} - \psi_\ell \chi'_\ell  \,.
\end{align}
Choosing~$\chi_\ell$ to be the regular and irregular solution gives, respectively, 
\begin{align}
\nonumber
Y_+ [\psi^\mathrm{reg}, \psi] &= {\cal W}[\psi^\mathrm{reg}_\ell, \psi_\ell] \,;\\
Y_- [\psi^\mathrm{irreg}, \psi] & = {\cal W}[\psi^\mathrm{irreg}_\ell, \psi_\ell] \,,
\end{align}
where~${\cal W}$ is the Wronskian. In other words, this symmetry is associated to linear Wronskians of the solutions, satisfying the Heisenberg algebra
\be
\{Y_+, Y_-\} = {\cal W}[\psi^\mathrm{reg}_\ell,\psi^\mathrm{irreg}_\ell] \equiv {\cal W} \,.
\label{Heisenberg comm}
\ee
From the definition of the Noether charge in Eq.~\eqref{eq:noether_chargeY}, it is clear that $Y_\pm$ and their Poisson bracket are radial functions, with the latter determined by the Wronskian of the independent solutions.

The system is also invariant under a special subset of conformal reparametrizations of the form
\begin{align}
z &\to \tilde{z} = g(z)\,; \nonumber \\
\psi_\ell (z) &\to \tilde{\psi}_\ell (\tilde{z}) = \sqrt{g'(z)} \psi_\ell(z)\, .
\end{align} 
In other words,~$\psi_\ell(z)$ transforms as a primary field with conformal weight~$1/2$~\cite{BenAchour:2022uqo}. The function~$g(z)$ is explicitly constrained to satisfy the equation
\begin{equation}
\label{Schw}
{\rm Sch}[g] \equiv \frac{g'''}{g'} - \frac{3}{2}\left(\frac{g''}{g'}\right)^2 =  2\mathcal{V}_\ell - 2(g')^2 (\mathcal{V}_\ell \circ g)\,,
\end{equation}
in terms of the Schwarzian derivative~${\rm Sch}[g]$ and~$\psi_\ell \circ g = \sqrt{g'} \psi_\ell$. This transformation represents the action of the Virasoro group on Sturm-Liouville operators~\cite{2006math.ph...2009O}, with solutions to Eq.~\eqref{Schw} forming a SL$(2,\mathbb{R})$ group. 

An alternative realization of this symmetry occurs through a ``conformal bridge'', by introducing a ``trivializing'' coordinate~$F_\ell$~\cite{BenAchour:2022uqo}:
\begin{align}
    z &\to F_\ell(z) = \frac{\psi^\mathrm{irreg}_\ell(z)}{\psi^\mathrm{reg}_\ell(z)}\,,\nonumber\\
    \psi_\ell(z) &\to \Phi_\ell(F_\ell) = \sqrt{\frac{{\rm d} F_\ell}{{\rm d} z}}\psi_\ell(z)\,.
\label{trivial and Phi}
\end{align}
This maps the original Sturm-Liouville problem~\eqref{eq:eomz} to the equation of motion for a free particle,
\be
\frac{{\rm d}^2 \Phi_\ell(F_\ell)}{{\rm d}F^2_\ell} = 0\,.
\ee
The free equation is invariant under {\it M\"obius transformations}, which comprise translations, inversions, and dilations of the trivializing coordinate~$F_\ell$:
\begin{align}
    F_\ell &\to M_\ell(F_\ell) = \frac{a F_\ell + b }{cF_\ell + d}\,,\quad \text{where } \, ad-bc =1\,;\nonumber\\
    \Phi_\ell (F_\ell) &\to \tilde \Phi_\ell (M_\ell(F_\ell))  = \sqrt{\frac{{\rm d} M_\ell(F_\ell)}{{\rm d} F_\ell}}\,\Phi_\ell\,,
\label{trivializing coord}
\end{align}
and which realize a hidden SL$(2,\mathbb{R})$ symmetry [in analogy to Eq.~\eqref{Schw}]. 
The corresponding Noether charges associated to this group can be related to the squared Wronskians among the solutions~\cite{BenAchour:2022uqo}
\begin{align}
\nonumber
Q_+ [\psi^\mathrm{reg}_\ell, \psi_\ell] & = \frac{1}{2}{\cal W}[\psi^\mathrm{reg}_\ell, \psi_\ell]^{\,2}\,; \\
\nonumber
Q_- [\psi^\mathrm{irreg}_\ell, \psi_\ell] & = \frac{1}{2}{\cal W}[\psi^\mathrm{irreg}_\ell, \psi_\ell]^{\,2}\,; \\
Q_0 [\psi^\mathrm{reg}_\ell \psi^\mathrm{irreg}_\ell, \psi_\ell] & = \frac{1}{2}{\cal W}[\psi^\mathrm{reg}_\ell, \psi_\ell]\, {\cal W}[\psi^\mathrm{irreg}_\ell, \psi_\ell] \,.
\end{align}
These satisfy the commutation relations
\begin{align}
& \{Q_+, Q_-\} = 2 {\cal W} Q_0\,; \quad \{Q_0, Q_\pm\} = \mp {\cal W} Q_\pm\,;  \nonumber \\
& \{Q_0, Y_\pm\} = \mp \frac{{\cal W}}{2} Y_\pm\,; \quad \{Q_\pm, Y_\mp\} = \pm  {\cal W} Y_\pm \,,
\end{align}
which, together with the Heisenberg algebra~\eqref{Heisenberg comm}, form a one-dimensional Schr\"odinger algebra, with central charge given by the conserved Wronskian~${\cal W}$.

This symmetry group is responsible for the ladder structure found for black holes. In particular, the generators of this group
\begin{align}\label{eq:generators}
    \delta_{Q_+}&= -\partial_{F_\ell}\,; \nonumber \\
    \delta_{Q_-}&=-F^2_\ell\partial_{F_\ell}+F_\ell\,; \nonumber \\ \delta_{Q_0}&=-F_\ell\partial_{F_\ell}+\frac{1}{2}\,,
\end{align}
satisfy the algebra
\begin{align}
[\delta_{Q_+}, \delta_{Q_-}] = - 2 \delta_{Q_0}\,, \qquad [\delta_{Q_0},\delta_{Q_\pm}] = \pm \delta_{Q_\pm}\,.
\label{Q algebra}
\end{align}
The quadratic Casimir operator,
\begin{align}
    \mathcal{C}_2 = \delta_{Q_0}^2 -\frac{1}{2}(\delta_{Q_+}\delta_{Q_-} + \delta_{Q_-}\delta_{Q_+})\,, \quad C_2 \psi_\ell = \frac{3}{4} \psi_\ell\,,
\end{align}
maps solutions of the equation of motion to other solutions. As shown in the next subsection, specialized to ABHs, the only relevant generator that is regular at the horizon is~$\delta_{Q_+}$, which generates translations along~$F_\ell$.

\subsection{M\"obius transformations for analog black holes}
\noindent
To apply the above methods to our problem, let us go back to the equation of motion~$H_\ell R_\ell = 0$ for the radial mode function,
given by Eqs.~\eqref{EOM-ABH 0} and~\eqref{EOM-ABH}. We perform the coordinate and field redefinitions
\begin{align}
\frac{{\rm d} z}{{\rm d}r} = r^{-3}\Delta^{-1}(r) \,; \quad \psi_\ell(r)\equiv r^{-3/2}R_\ell(r)\,.
\label{r to z us}
\end{align}
This removes the ``friction'' term from the equation and gives Eq.~\eqref{eq:eomz} with
\begin{align}
\label{calV}
    \mathcal{V}_\ell (z(r)) & =  - \frac{r_+^8 e^{4 r_+^4 z} \left[4 \left(\ell^2+\ell-6\right)+9 e^{4 r_+^4 z}\right]}{4 \left(e^{4 r_+^4 z}-1\right)^2} \nonumber \\
    & = - \frac{1}{4} \left(r^4-r_+^4\right) \Big[(2 \ell-3) (2 \ell+5) r^4-9 r_+^4\Big]\,.
\end{align}
Starting with the effective potential, one can solve the equation of motion and obtain hypergeometric solutions, in analogy to our discussion in the previous Section. Explicitly, the regular and irregular solutions are given by
\begin{align}
\nonumber
\psi^\mathrm{reg}_\ell(r) & = \left(\frac{r}{r_+}\right)^{\ell-\frac{3}{2}} {}_2 F_1\left(\frac{3-\ell}{4},-\frac{\ell}{4},1; 1 - \frac{r_+^4}{r^4}\right) \,;\\
\psi^\mathrm{irreg}_\ell(r) & =  \left(\frac{r}{r_+}\right)^{-\ell-\frac{5}{2}} \, _2F_1\left(\frac{\ell+1}{4},\frac{\ell+4}{4},\frac{2\ell+5}{4} ;\frac{r_+^4}{r^4}\right)
\,.
\end{align}
It can be easily shown that the Wronskian of the two solutions reads 
\be
{\cal W} = {\cal W}[\psi^\mathrm{reg}_\ell,\psi^\mathrm{irreg}_\ell] = c_\ell r_+^4 \,,
\label{Wronskian us}
\ee
where the derivatives in the Wronskian are with respect to $z$. Furthermore, we see that the Wronskian is set by the ABH horizon, up to an~$\ell$-dependent prefactor~$c_\ell$.
This feature echoes the known result that diffeomorphisms that preserve the near-horizon geometry are associated to a Virasoro algebra, with central charge dictating the black hole entropy and Hawking temperature, containing a SL($2,\mathbb{R}$) subgroup~\cite{Govindarajan:2000ag,Birmingham:2001qa, Carlip:1998wz, Cadoni:2005ej, Hotta:2000gx, Carlip:2011vr, Donnay:2016ejv, Akhmedov:2017ftb} (see brief discussion in the next subsection).

The trivializing coordinate~\eqref{trivializing coord}, which maps the equation of motion to that of a free field, is given by
\be
F_\ell = \frac{\psi^\mathrm{irreg}_\ell}{\psi^\mathrm{reg}_\ell} = \left( \frac{r_+}{r} \right)^{2\ell+1} \frac{_2F_1\left(\frac{\ell+1}{4},\frac{\ell+4}{4},\frac{2\ell+5}{4} ; \frac{r_+^4}{r^4}\right)}{{}_2 F_1\left(\frac{3-\ell}{4},-\frac{\ell}{4},1; 1 - \frac{r_+^4}{r^4}\right)}\,.
\ee
Since~$F_\ell \to - \infty$ as~$r \to r_+$, it is immediately clear that not all of the symmetry generators in Eq.~\eqref{eq:generators}
are preserved. Indeed, only the translation generator~$\delta_{Q_+}=-\partial_{F_\ell}$, which is regular at the horizon, is well defined.
In other words, the presence of a horizon induces spontaneous symmetry breaking of the SL$(2,\mathbb{R})$ group down to the Abelian subgroup of translations~\cite{BenAchour:2022uqo}. (Notice the similarity of this translation subgroup with the one associated to the closed CKV of Sec.~\ref{sec: CCKV-ABH}.)

Using~\eqref{Wronskian us}, the trivializing coordinate satisfies
\be
\frac{{\rm d} F_{\ell}}{{\rm d}z} =  \frac{c_\ell\, r_+^4}{\big(\psi^{\mathrm{reg}}_\ell\big)^2}\,.
\ee
Substituting this, together with~\eqref{r to z us}, the free field variable~$\Phi_\ell$ in Eq.~\eqref{trivial and Phi} is given by
\be
\Phi_\ell(r) =\sqrt{c_\ell}\,r_+^2  \frac{\psi_\ell}{\psi^{\mathrm{reg}}_\ell} \,.
\ee
The action of the translation generator on this field is given by
\begin{align}
    \delta_{Q_+}\Phi_\ell &=\frac{r^3 \Delta(r)}{r_+^2\sqrt{c_\ell}} \Big(\psi_\ell \partial_r \psi^\mathrm{reg}_\ell-\psi_\ell^\mathrm{reg} \partial_r \psi_\ell \Big)\,.
\end{align}
This shows that the action of such generator annihilates any regular solution to the problem~\eqref{Rcan}. In particular, when focusing on the ground state levels~$\ell = 0, 3$, it is possible to show that this generator reproduces the known ladder operator of Eq.~\eqref{ladderABH} as
\begin{align}
    \delta_{Q_+}\Phi_\ell &\sim D^-_\ell R_\ell\,,
\label{del Q our case}
\end{align}
where this relation holds up to a constant for $\ell=3$ and up to an overall factor of $r^{3}$ for $\ell=0$. (Notice the decomposition of~$\psi_\ell$ in terms of the radial profile~$R_\ell$.) Let us stress that, once the lowering ladder operator~$D^-_\ell$ is recognized from the M\"obius transformations, the raising  operator~$D^+_\ell$ can be obtained through the Hamiltonian decomposition via Eq.~\eqref{HDD}. Similarly, the relations among the big and small ladder operators~$D^\pm_\ell$ and~$\tilde{D}^\pm_\ell$ shown in the previous Section allow us to deduce the latter from the M\"obius symmetry. Following Refs.~\cite{Berens:2022ebl, Combaluzier-Szteinsznaider:2024sgb}, it is then possible to obtain higher~$\ell$ operators using the relation
\begin{align}
\delta^\ell_{Q_+} \Phi_\ell\sim D^+_{\ell-4} \cdots D^+_{0} \delta_{Q_+}  \cdots D^-_{\ell}R_\ell\,, \\
\text{for} \quad \ell = 0 + 4n\,, \,\,\, \ell = 3 + 4n\,, \,\, (n \in \mathbb{N})\,.\nonumber 
\end{align}
In other words, the existence of the ladder operator hints toward the existence of the hidden M\"obius transformations associated to spherically symmetric space-times, and therefore holds for both black holes in GR and analog gravity.

\subsection{Near-horizon and asymptotic regions of analog black holes}
\label{near-horizon}
\noindent
The existence of a Schr\"odinger symmetry group associated to static perturbations on black hole backgrounds can also be made transparent by studying the near-horizon~\cite{Govindarajan:2000ag,Birmingham:2001qa,Gupta:2001bg,Bertini:2011ga} and asymptotic regions of these objects, which are crucial to establishing the ingoing boundary condition and the determination of TLNs, respectively.

Let us first study the near-horizon limit,
\be
x = r - r_+ \ll r_+\,. 
\ee
For this purpose, we can go back to the radial wave function equation~\eqref{eq:canonical_radial_eom}, restoring the  dependence on the perturbation frequency $\omega$. Using~$\Delta(r) \simeq 4r_+ x$ in this limit, and performing the field redefinition~$\hat{\psi}(r) = \sqrt{x} R_\ell$, it is easy to see that~\eqref{eq:canonical_radial_eom} becomes, to leading order in~$1/x$, 
\begin{equation}
\label{CFT}
\left[\frac{{\rm d}^2}{{\rm d} x^2} + \frac{\lambda_\text{\tiny CFT}}{x^2} \right] \hat{\psi}(x) = 0\,;\quad \lambda_\text{\tiny CFT} = \frac{1}{4} \left(1 + \frac{\omega^2r_+^2}{4}\right)\,.
\end{equation}
 Notice that the angular variables have effectively decoupled in the near-horizon limit, with the multipole moment~$\ell$ entering at next-to-leading order corrections in~\eqref{CFT}, leaving only a one-dimensional problem along the radial~$x$ direction.\footnote{To be precise, the angular terms give a correction of the form~$\frac{-4\ell(\ell+1)+4+5r_+^2 \omega^2}{16r_+x} \hat{\psi}$ to Eq.~\eqref{CFT}, which is subleading for~$x\ll 1$.} Equation~\eqref{CFT} is recognized as a one-dimensional, non-relativistic conformal field theory~(CFT), known as conformal quantum mechanics, with conformal weight~$\lambda_\text{\tiny CFT}$~\cite{Camblong:2003mb, Camblong:2003mz,Camblong:2004ye}. Let us stress that this result holds for black holes both in GR and in analog gravity~\cite{Govindarajan:2000ag,Birmingham:2001qa, Carlip:1998wz, Cadoni:2005ej, Hotta:2000gx, Carlip:2011vr, Donnay:2016ejv, Akhmedov:2017ftb}. Its universality stems from the fact that $\Delta$ vanishes linearly,~$\Delta(r) \sim r-r_+$, as~$r \rightarrow r_+$.

We can define the corresponding classically scale invariant De Alfaro-Fubini-Furlan Hamiltonian~\cite{deAlfaro:1976vlx}
\begin{equation}
H = - \frac{{\rm d}^2}{{\rm d} x^2} - \frac{\lambda_\text{\tiny CFT}}{x^2} = p^2 - \frac{\lambda_\text{\tiny CFT}}{x^2}\,, 
\end{equation}
with momentum~$p = -{\rm i} \frac{{\rm d}}{{\rm d} x}$.\footnote{See~\cite{Essin:2006sic} for a pedagogical discussion of quantum mechanics with~$1/x^2$ potential.}  This system is invariant under translations generated by~$H$, dilations~$D$, and special conformal transformation~$K$~\cite{Jackiw:1972cb}
\begin{align}
D &= \frac{\rm i}{4} \left(x \frac{{\rm d}}{{\rm d} x} + \frac{{\rm d}}{{\rm d} x} x \right) =   -  \frac{1}{4} (x p + p x)\,; \nonumber \\
K &= 
\frac{1}{4}x^2\,,
\end{align}
where we have restricted the action of these generators on time-independent (static) mode functions~$\hat{\psi}(x)$. 
These satisfy a~$\mathfrak{sl}(2,\mathbb{R})$ algebra~\cite{osti_4271130}
\begin{align}
[D,H] = - {\rm i} H\,; \quad [D,K] = {\rm i}K\,; \quad [H,K] = 2 {\rm i} D\,,
\end{align}
with quadratic Casimir operator~$\mathcal{C}_2 \hat{\psi}=[D^2 - \frac{1}{2}(HK + KH)]\hat{\psi} =  \frac{4\lambda_\text{\tiny CFT}+3}{16} \hat{\psi}$\,.
This is the familiar Schr\"odinger algebra, as expected from the scaling invariance in non-relativistic CFTs~\cite{Horvathy:2009kz}.
The existence of this Schr\"odinger symmetry matches the understanding of the previous subsection regarding the properties of the equations of motion on ABH space-times and ladder operators. 

We can recover similar results by looking at asymptotic spatial infinity,~$r\gg r_+$. By performing the field redefinition~$\tilde{\psi}_\ell= \sqrt{\Delta} R_\ell$, the resulting static wave equation reads 
\begin{equation}
\left[\frac{{\rm d}^2 }{{\rm d} r^2} - \frac{\ell(\ell+1)}{r^2}\right] \tilde{\psi}_\ell(r) = 0\,.
\end{equation}
This equation is recovered exactly when considering the~$\mathrm{AdS}_2$ metric of Eq.~\eqref{metricCCKV}, {\it i.e.},
\begin{equation}
\Big[\Box_{\mathrm{AdS}_2}-\ell(\ell+1)\Big] \tilde{\psi}_\ell(r)= 0\,,
\end{equation}
in terms of an effective mass term~$m_\ell^2 = \ell (\ell+1)$, as we discussed in Sec.~\ref{sec: CCKV-ABH}. In other words, the equation that dominates the dynamics at large~$r$ of a massless scalar in the ABH background is identical to the equation of a massive scalar on an~$\mathrm{AdS}_2$ space-time. As such, it enjoys the usual invariance under the~$\mathrm{AdS}_2$ isometries, namely dilations, time translations and conformal inversions, which reproduce exactly the SL$(2,\mathbb{R})$ symmetry group discussed in the previous subsection (see also Sec.~\ref{sec: Near-zone}).

Therefore, we have uncovered the nature of the SL$(2,\mathbb{R})$ symmetry group responsible for M\"obius transformations as associated to symmetries which are manifest both in the near-horizon and asymptotic regions, where the crucial information to compute TLNs are determined. We leave to future work a deeper investigation of this property (though see Ref.~\cite{Kehagias:2025tqi} for recent developments along this direction in the context of black hole perturbation theory). 


\section{Darboux transformations and Supersymmetry}
\label{sec: susy}
\noindent
We have seen that the ladder structure that enforces the vanishing of TLNs can be related to M\"obius symmetry transformations acting on a static scalar field, giving rise to a SL$(2,\mathbb{R})$ group. In this Section we will show that there is another symmetry perspective which explains the existence of such ladder operators, arising from Darboux transformations and non-relativistic supersymmetric quantum mechanics (SUSY QM). (See~\cite{Cooper:1994eh} for a review of SUSY QM.)

\subsection{Darboux transformations and SUSY QM}
\noindent
Let us start by introducing {\it intertwining operators}~\cite{Anderson:1991kx}. These operators~$\mathcal{D}$ change a differential operator~$H$ into another~$\bar{H}$ according to the relation
\begin{equation}
\label{interwinings}
\mathcal{D} H = \bar{H} \mathcal{D}\,.
\end{equation}
Thus, any eigenfunction~$\psi$ of~$H$, satisfying~$H \psi = \lambda \psi$, corresponds to an eigenfunction~$\bar{\psi} = \mathcal{D} \psi$ of~$\bar{H}$ with the same eigenvalue. In other words, the two operators~$H$ and~$\bar{H}$ are isospectral (except for the ground state of~$H$, which is unpaired). In the following, we will mostly be interested in intertwining operators involving only first derivatives in the time/spatial variable~$z$. At this point, the reader can already identify the ladder operators for TLNs as belonging to the class of intertwining operators, as can be readily seen from Eqs.~\eqref{HD-BH} and~\eqref{HD-ABH}, and as we will show in detail in the following. 

When considering operators of the Schr\"odinger type
\begin{equation}
H = -\partial_z^2 + V(z)\,; \qquad  \bar{H} = -\partial_z^2 + \bar{V}(z)\,,
\end{equation}
the intertwining operator~$\mathcal{D} = \partial_z + W(z)$ must satisfy the relation
\begin{equation}
\bar{V}(z) = V(z) + 2 \partial_z W(z)\,,
\end{equation}
where the generating function~$W(z)$ solves the Riccati equation
\begin{equation}
\partial_z W - W^2 + V = \mathcal{C}\,,
\end{equation}
in terms of an arbitrary constant~$\mathcal{C}$.  Alternatively, the generating function can be derived through the condition~$W(z)= - \partial_z\log \psi^\mathrm{gs} (z)$, in terms of ground-state solutions~$\psi^\mathrm{gs}$ of the original problem.  By combining these equations, one can constrain the form of~$W$ in terms of the associated potentials~$V$ and~$\bar{V}$ to be
\begin{equation}
\label{superpotential}
W(z) = \frac{\partial_z (V + \bar{V})}{2(\bar{V} - V)}\,,
\end{equation}
solving the second-order non-linear equation~\cite{Lenzi:2021wpc, Lenzi:2021njy}
\begin{equation}
\label{DarbouxV}
\partial_z \left(\frac{\partial_z (V - \bar{V})}{V - \bar{V}}  \right) + 2\partial_z \left(\frac{\partial_z \bar{V}}{V - \bar{V}}  \right) - (V - \bar{V}) = 0\,.
\end{equation}
Transformations relating the pairs~$(\psi, V)$ and~$(\bar{\psi},\bar{V})$ are dubbed {\it Darboux transformations}~\cite{1999physics...8003D}. They provide an elegant way to produce new exactly solvable potentials from an initial solvable potential. This infinite set of master equations/potentials is referred to as Darboux branch, with Eq.~\eqref{DarbouxV} providing the condition that any potential in the branch must satisfy (notice that swapping the sign of $W$ amounts to consider the reverse transformation from $\bar{V}$ to $V$)~\cite{Glampedakis:2017rar,Lenzi:2021njy}.  

It has been shown that any spherically symmetric space-time with a warped geometry---which allows one to decompose the metric perturbations in spherical harmonics and to decouple modes with different harmonic number~$(\ell,m)$ and parity---possesses a hidden symmetry in the perturbations, called Darboux covariance~\cite{Glampedakis:2017rar, Lenzi:2021njy, Lenzi:2021wpc}. This covariance is associated to the infinite set of master equations generated through these transformations. (For example, it has been proven equivalent to the Chandrasekhar transformations between parity even and odd perturbations on black hole space-times~\cite{Chandrasekhar:1975zza, Chandrasekhar:1979iz}).

This property has been developed in the context of Sturm-Liouville problems (as such, it is constrained by the number of regular and irregular singularities of the effective potentials~\cite{Anderson:1991kx}), and is reminiscent of SUSY QM~\cite{Cooper:1994eh}, where the quantum description of systems with double degeneracy of energy levels is realized. 
This duality is realized through the introduction of ladder operators,~$A$ and~$A^\dagger$, which allow us to rewrite the Hamiltonian in the form
\begin{align}
\label{SUSY}
    H=A^\dagger A
    \,, \qquad \begin{cases}
    A=\partial_z+W(z) \\
    A^\dagger=-\partial_z+W(z)\,,
    \end{cases}
\end{align}
by introducing the superpotential~$W(z)$. Similarly, its 
supersymmetric partners satisfy~$\bar{H} \bar{\psi}=0$, with~$\bar{H}= A A^\dagger$ and~$\bar{\psi}=A \psi$. 
This system can be explicitly written as an anticommutator of supercharges~$Q$ and~$Q^\dagger$~\cite{Cooper:1994eh}
\begin{align}
    &\mathcal{H} = \{Q,Q^\dagger\}=\begin{bmatrix}
    H & 0 \\
    0 & \bar{H} 
    \end{bmatrix}\,, \nonumber \\
    &Q = \begin{bmatrix}
    0 & 0 \\
    A & 0 
    \end{bmatrix}\,, \qquad Q^\dagger = \begin{bmatrix}
    0 & A^\dagger \\
    0 & 0 
    \end{bmatrix}\,,  \nonumber \\
    & \Psi  = 
    \begin{bmatrix}
    \psi \\
    0
    \end{bmatrix}\,, \qquad
    \bar{\Psi}  = 
    \begin{bmatrix}
    0 \\
    A \,\psi 
    \end{bmatrix}.
\end{align}
This provides the construction of SUSY QM.

In the following, we will show that this construction applies straightforwardly to ABHs, and provides another way to derive the ladder operators.

\subsection{Application to  analog black holes}
\noindent
We begin by recognizing that the equation of motion \eqref{eq:eomz} can be recast, for each multipole~$\ell$, as a Sturm-Liouville problem of the form
(setting $V_\ell \equiv - \mathcal{V}_\ell$)
\begin{align}
\label{HDarbouxABH}
    H_\ell\psi_\ell(z) = \left[-\frac{{\rm d}^2}{{\rm d} z^2} + V_\ell \right] \psi_\ell(z)=0\,.
\end{align}
This admits regular and irregular homogeneous solutions, as shown in Eq.~\eqref{homo}.

In order to determine which solutions/states are effectively connected through the Darboux/SUSY transformations, let us consider the master equation~\eqref{DarbouxV} and take~$V = V_\ell$ and~$\bar{V} = V_{\ell + n}$. Then, using the explicit expression~\eqref{calV} for the effective potential, the master equation~\eqref{DarbouxV} gives the condition
\begin{equation}
(n-4) (n+4) (2 \ell+n-3) (2 \ell+n+5) = 0\,,
\end{equation}
which admits, as solution for the ladder step size,
\begin{equation}
n  = \pm 4\,, \qquad  n = 3 - 2\ell\,, \qquad n = - 5 - 2 \ell\,. 
\end{equation}
In other words, considering the first two sets of solutions for~$n$, we recover the step sizes of the big and small ladder operators~$\ell \to \ell \pm 4$ and~$\ell \to 3- \ell$, respectively. (Notice that the last solution for~$n$ necessarily yields unphysical solutions with~$\ell < 0$.)  At this point, we can evaluate the superpotential of Eq.~\eqref{superpotential} as (expressing~$z$ in terms of~$r$)
\begin{align}
\label{Wbigsmall}
 &W_{n = 4} = 
\frac{(2 \ell+5)^2 r^4-(2 \ell (\ell+5)+17) r_+^4}{4 \ell+10} \,; \nonumber \\
&  W_{n = -4} = W_{n = 3-2\ell} = \frac{-(3-2 \ell)^2 r^4 + (2 (\ell-3) \ell+9) r_+^4}{4 \ell-6}\,.
\end{align}
Using Eq.~\eqref{SUSY} and the explicit forms of the superpotentials $W$, it is now possible to recover the big and small ladder operators as 
\begin{align}
{\rm big \,\, ladder :}  \qquad 
&\begin{cases}
    A_\ell^{\mathrm{big}} \sim D^-_{\ell}   \\
    A_\ell^{\mathrm{big} \, \dagger} \sim  D^+_{\ell-4}  
\end{cases} \\
 {\rm small \,\, ladder :}  \qquad 
&\begin{cases}
    A_\ell^{\mathrm{small}}   \sim \tilde D^+_{\ell}   \\
    A_\ell^{\mathrm{small} \, \dagger}  \sim \tilde D^-_{\ell} \,.
\end{cases}
\end{align}
This shows that the ladder symmetries, responsible for carrying the information about vanishing TLNs, arise from the Darboux transformations associated to the Sturm-Liouville problem. Let us stress again that such conclusions are general and valid for any spherically symmetric warped geometry, thus including also black holes in GR.

From the explicit form of the
superpotential in Eq.~\eqref{Wbigsmall}, one can directly extract the ground state solution using, for example,~$W_{n = -4}(z) = -\partial_z\log \psi_\ell^\mathrm{gs}(z)$, thus getting\footnote{Note that one could also use the superpotential associated with the transition $\ell \to \ell + 4$, namely $W_{n = 4}$, in which case one obtains $\psi_{\ell+4}^\mathrm{gs}$.}
\begin{align}
\psi_\ell^\mathrm{gs} &= \exp\left[- \int {\rm d}z \, W_{n = -4}(z)\right] \nonumber \\
& = c_1 \left(\frac{r_+}{r}\right)^\frac{3-2\ell}{2} \left(1-\frac{r_+^4}{r^4}\right)^{\frac{\ell(\ell-3) }{4(2\ell-3)}}\,.
\end{align}
As expected from the construction of SUSY-QM, such state is properly annihilated by the operator $A_\ell^{\mathrm{big}} \sim \partial_z + W_{n = -4}$, {\it i.e.}, $A_\ell^{\mathrm{big}} \psi_\ell^\mathrm{gs} = 0$, indicating that it represents the lowest energy state of each $\ell$-spectrum,  with associated eigenvalue obtained as
\begin{equation}
H_\ell \psi_\ell^\mathrm{gs} = - \frac{\ell^2  (\ell-3)^2 }{(2 \ell-3)^2} r_+^8 \psi_\ell^\mathrm{gs}\,.
\end{equation}
In particular, note that at each level of the ladder, the ground state corresponds to the homogeneous solution that is regular at the horizon, rather than the irregular one. This behavior mirrors the familiar pattern of unbroken SUSY~\cite{Cooper:1994eh}.

Before concluding this Section, let us comment about the extension of these results to time-dependent perturbations. Indeed, it has been shown that 
another interesting feature associated to the existence of intertwining operators and Darboux covariance is the property of isospectrality of  paired time-dependent potentials, {\it i.e.}, the fact that along the Darboux branch the reflection and transmissions coefficients, as well as the energy levels, are the same~\cite{Glampedakis:2017rar, Lenzi:2021njy}. This property has been used to provide a natural explanation for the identical quasinormal modes spectra and scattering amplitudes shared between certain black hole potentials, such as the parity even and parity odd sectors of perturbations on Schwarzschild black holes~\cite{Chandrasekhar:1975zza, Chandrasekhar:1979iz}.
In the case of canonical ABHs, when reintroducing a nonzero frequency~$\omega$, the equation of motion~\eqref{HDarbouxABH} takes the form [we stress that the coordinates relation~$z = z(r)$ is known]
\begin{equation}
 \left[-\frac{{\rm d}^2}{{\rm d} z^2} +V_\ell - \frac{r^{10} \omega^2}{c_s^2} \right] \psi_\ell(z)=0\,.
\end{equation}
The introduction of the frequency term shifts the effective potential and can be proven to  invalidate the Darboux master equation~\eqref{DarbouxV}. In other words, the Darboux covariance associated to ABHs is strictly enforced only for static perturbations.

\section{Near-zone  and asymptotic~${\rm AdS}_2$}
\label{sec: Near-zone}
\noindent
Before concluding, let us continue the discussion on time-dependent perturbations.
As briefly discussed in Sec.~\ref{sec: near zone symmetries}, the  near-zone approximation~$\omega (r-r_+)\ll1$ for 4D Schwarzschild black holes allows one to express the wave equation in the static limit~$\omega \rightarrow 0$ in terms of a~$\mathrm{AdS_2\times S^2}$ effective metric~\cite{Hui:2022vbh}. In the following, we will argue that a similar scenario arises for canonical~$3+1$ ABHs where, in the near-zone approximation, one recovers the aforementioned Bertotti-Robinson space-time only in the asymptotic~$r \gg  r_+$ limit.

Let us begin with the metric of Eq.~\eqref{eq:canonical_metric}. At finite frequency~$\omega$, the wave equation picks up an additional term, which can be approximated in the near-zone regime according to 
\begin{align}\label{eq:nz_split}
    \frac{r^4\omega^2}{\Delta(r)} = \left(\frac{r_+^4}{\Delta(r)}+r^2\right) \omega^2 \simeq \frac{r_+^4}{\Delta(r)} \omega^2\,.
\end{align}
In other words, the near-zone approximation consists of neglecting the second term, given by~$r^2\omega^2$, while keeping the linear nature of the pole at~$r_+$.\footnote{Usually, this is done to simplify the irregular singularity at spatial infinity into a regular one.}  As is well known (see, {\it e.g.},~\cite{Charalambous:2022rre}), there is no unique way to decompose the~$\frac{r^4\omega^2}{\Delta(r)}$ term in the wave equation, or in other words, no unique near-zone split. We will see that the above split is the choice needed to obtain~$\mathrm{AdS_2\times S^2}$ asymptotically. 

With the replacement~\eqref{eq:nz_split}, the radial equation of motion then takes the form
\begin{align}
\label{eq:canonical_radial_eom_nz}
    \Big(\Delta(r)R_{\ell }'(r)\Big)' + \frac{r_+^4}{\Delta(r)}\omega^2 R_{\ell }-\ell(\ell+1) R_{\ell}(r)=0\,.
\end{align}
This is identical to the equation of a massless scalar on the effective metric
\begin{equation}
\label{eq:nz_metric_ABH}
{\rm d}s^2 = - \frac{\Delta(r)}{r_+^2}  \mathrm{d}t^2 + \frac{r_+^2}{\Delta(r)} \mathrm{d}r^2+r_+^2 \mathrm{d}\Omega_{2}\,.
\end{equation}
We can rewrite this in terms of a new variable
\begin{align}
    \xi = \frac{1}{2}\cosh^{-1} \left(\frac{r^2}{r_+^2} \right)\,,
\end{align}
and, upon appropriate constant rescaling of the time coordinate, the metric~\eqref{eq:nz_metric_ABH} becomes
\begin{align}
\label{eq:nz_metric_ABH_xi}
    {\rm d}s^2 = r_+^2\left(-\frac{2\sinh^2 \xi}{1+ \tanh^2\xi}{\rm d}\tau^2 + {\rm d}\xi^2 + {\rm d}\Omega_{2}\right)\,.
\end{align}
Contrary to Schwarzschild black holes, Eq.~\eqref{eq:nz_metric_ABH_xi} is not exactly~$\mathrm{AdS_2\times S^2}$ space-time at any radial distance~$\xi(r)$.
Nevertheless, in the matching zone~$r_+\ll r\ll \omega^{-1}$, we have~$\tanh\xi \simeq 1$, and the near-zone metric approaches~$\mathrm{AdS_2\times S^2}$. 
In other words,~$3+1$ canonical ABH space-times can be mapped into a Bertotti-Robinson space-time only in the asymptotic region~$r \gg r_+$.  

Let us now obtain the $\mathfrak{sl}(2,\mathbb{R})$ generators of this space-time and its CKVs, as was done for 4D Schwarzschild black holes. Performing the coordinate transformation~$\xi (r) \to \rho (r)$ in the asymptotic~$r \gg r_+$ region, with
\begin{align}
\label{rhor}
    \rho(r) = \frac{1}{4}\left[2 + \sqrt{2\left(\frac{r^2}{r_+^2}+1\right)}\right] \simeq \frac{\sqrt{2}}{4}\frac{r}{r_+}\,,
\end{align}
the~$\mathrm{AdS_2 \times S^2}$ metric becomes
\begin{align}\label{eq:schwarzschild_nz_metric_rho}
    {\rm d} s^2 = - \rho(\rho-1) {\rm d}t^2 + \frac{{\rm d}\rho^2}{\rho(\rho-1)} + {\rm d}\Omega_{2}\,,
\end{align}
where~$ \rho(\rho-1) = \frac{1}{8} \left( \frac{r^2}{r_+^2}-1 \right)$. This reproduces the near-zone metric for 4D Schwarzschild black holes~\cite{Hui:2022vbh, Charalambous:2022rre}. Therefore, when put into this form, the generators and CKVs of canonical ABHs in the asymptotic near-zone region are a
coordinate transformation away from Eq.~\eqref{metricCCKV}.

Following the Love symmetry results for Schwarzschild black holes~\cite{Charalambous:2022rre}, one can derive the~$\mathfrak{sl}(2,\mathbb{R})$ generators, which read [see also Eq.~\eqref{SL2RLove}]
\begin{subequations}
    \begin{align}
    T&=-\frac{r_+}{2}\partial_t\,; \\
    L_{-1} &= e^{2t/r_+}\left(-\frac{\sqrt{r^4-r_+^4}}{r}\partial_r + \frac{r_+}{2}\sqrt{\frac{r^2+r_+^2}{r^2-r_+^2}}\partial_t\right)\,; \\
    L_{+1} &= e^{-2t/r_+}\left(\frac{\sqrt{r^4-r_+^4}}{r}\partial_r + \frac{r_+}{2}\sqrt{\frac{r^2+r_+^2}{r^2-r_+^2}}\partial_t\right)\,.
    \end{align}
\end{subequations}
From a coordinate transformation alone, we recover the equivalent inverse Hawking temperature of the ABH being that of a seven-dimensional Schwarzschild black hole, 
\begin{align}
    \beta = r_+/2\,,
\end{align}
as expected from the form of~$\Delta(r)/r^2 = 1-(r_+/r)^4$. The quadratic Casimir of these operators gives the wave operator that one would get from the metric in Eq.~\eqref{eq:nz_metric_ABH_xi} in the asymptotic limit~$r \gg r_+$.

Given that the static solution represents the highest weight representation of $\mathfrak{sl}(2,\mathbb{R})$, it should satisfy the condition~\cite{Charalambous:2022rre}
\begin{equation}
(L_{+1})^{\ell+1} R_\ell^{\text{\tiny nz-ABH},\, \tiny{r \gg r_+}} \propto \partial^{\ell+1}_\rho R_\ell^{\text{\tiny nz-ABH},\, \tiny{r \gg r_+}}
= 0\,,
\end{equation}
which enforces a polynomial behavior for~$R_\ell^{\text{\tiny nz-ABH}}$ in~$\rho$, and thus in~$r$ using Eq.~\eqref{rhor}. In other words, the asymptotic SL$(2,\mathbb{R})$ symmetry group restricts the static ABH solution at large distances to only positive powers in~$r$, with consequent absence of decaying modes. While this could hint toward the vanishing of TLNs for every~$\ell$, in  sharp contrast with the prediction obtained for the full ABH space-time (where TLNs were found to vanish only for~$\ell = 0 + 4n, 3 + 4n$ with~$n \in \mathbb{N}$), we caution the reader that this approach is valid only assuming the leading large distance term~$r \gg r_+$, in particular neglecting lower-order terms, from which TLNs would be read. Consequently, the reliability of any inferred vanishing TLN within this approximation is limited. On this basis, it appears that no Love symmetry is present in such systems.

\section{Conclusions}
\label{conclusions}
\noindent
The vanishing of TLNs for asymptotically flat black holes in four-dimensional GR is a striking and robust result, confirmed to all perturbative orders~\cite{Combaluzier-Szteinsznaider:2024sgb, Kehagias:2024rtz, Gounis:2024hcm,Parra-Martinez:2025bcu}. While this property does not hold universally---breaking down in higher dimensions or including environmental effects---it has motivated extensive efforts to uncover its origin. A leading explanation invokes hidden symmetries that emerge not as space-time isometries but at the level of the perturbation equations, including SL$(2,\mathbb{R})$ structures and ladder operators from coordinate invariance. Similar structures arise in effective geometries like~$\mathrm{AdS_2 \times S^2}$~\cite{Katagiri:2022vyz}, where CKVs help account for the observed behavior, suggesting that the vanishing of TLNs reflects deep symmetry principles in black hole physics.

These insights naturally raise the question of whether similar features appear in non-relativistic systems. Analog gravity models---especially ABHs---provide a promising test bed. While governed by fluid dynamics rather than Einstein’s equations, ABHs reproduce many kinematic features of black holes, such as horizons, quasinormal modes, and Hawking-like radiation~\cite{Visser:1997ux}. In this context, we have shown in earlier work~\cite{DeLuca:2024nih} that ABHs can exhibit vanishing TLNs for specific angular multipoles, a behavior that can be traced to ladder symmetries in their perturbation equations --- echoing the role of symmetry in the GR case and offering new insights into which aspects of black hole physics are truly universal.

In this paper, we have investigated this connection in detail and demonstrated that the ladder symmetries observed in analog gravity models can be traced to structural properties of the underlying wave equation, mirroring those found in general relativistic black hole space-times. First, in Sec.~\ref{sec: CCKV-ABH}, extending previous works~\cite{Cardoso:2017qmj, Cardoso:2017egd, Katagiri:2022vyz}, we provided a geometric understanding of ladder operators through the existence of closed CKVs of 1+1 dimensional effective space-times. In particular, there exist mass ladder operators that map solutions of the wave equation with a certain mass (or multipole moment) to solutions with a shifted mass. These operators coincide with those found when studying the perturbation equations in the usual four-dimensional space-time. These results are general and valid for any spherically symmetric space-time, thus including black holes in GR and analog gravity.

Then, in Sec.~\ref{sec: M\"obius}, following Ref.~\cite{BenAchour:2022uqo}, we showed that the same ladder operators may be explained through the existence of a set of symmetry transformations, referred to as M\"obius transformations, which comprise translations, dilations and special conformal transformations, and form a~$\mathfrak{sl}(2,\mathbb{R})$ algebra. The spontaneous symmetry breaking of this group due to the existence of the ABH horizon leaves an Abelian subgroup unbroken, whose generator coincides with the ladder operator. This symmetry group also appears when studying the near-horizon or asymptotic regions of ABH space-times, where the information for computing TLNs is obtained. 

In Sec.~\ref{sec: susy}, we proved, for the first time, that the ladder operators belong to the family of intertwining operators, usually associated in GR to Darboux transformations. These symmetries map features of a Sturm-Liouville problem, such as the wave function and potential of a Schr\"odinger-like equation, to another pair characterized by the same spectral features. We found that all levels of the ladder diagram are related by Darboux transformations, thus enforcing the common vanishing of TLNs. This structure resembles the properties of supersymmetric quantum mechanics, with each potential in Darboux pairs identified as partner potentials, and can be directly applied to the study of TLNs of BHs in GR. 

Finally, in Sec.~\ref{sec: Near-zone}, we found that only the asymptotic limit~$r \gg r_+$ of the near-zone region of~$3+1$ ABH metrics yields an~$\mathrm{AdS_2\times S^2}$ space-time, in contrast with Schwarzschild black holes in GR, where this identification holds at any distance. One can define ``asymptotic'' $\mathfrak{sl}(2,\mathbb{R})$ generators, whose quadratic Casimir matches the differential operator of the equation of motion, so that  asymptotic solutions fall under highest weight representations of $\mathfrak{sl}(2,\mathbb{R})$. However, since these properties only hold in the asymptotic region of ABH space-times, it stands to reason that no Love symmetry exists for such systems.

There are several directions for future research. An immediate goal is to generalize the computation of TLNs, as well as the existence of ladder operators, to spinning ABHs, in order to perform a detailed comparison with Kerr solutions in GR, where ladder operators have also been found~\cite{Hui:2022vbh}. Furthermore, it would be interesting to deepen our understanding of hidden symmetries for static and time-dependent perturbations on spherically symmetric space-times in terms  of integrability. 
Exploring these extensions is left to future work.

\vspace{0.3cm}
\section*{Acknowledgments}
\noindent
We thank M. Cvetic, J. Heckman, L. Hui, M. Ivanov,  C. Keeler, and A. Riotto for useful comments. We also thank J. Beltrán Jimenez, D. Bettoni and P. Brax for discussions in the early stages of the project.
V.D.L. is supported by NSF Grants No.~AST-2307146, No.~PHY-2513337, No.~PHY-090003, and No.~PHY-20043, by NASA Grant No.~21-ATP21-0010, by John Templeton Foundation Grant No.~62840, by the Simons Foundation [MPS-SIP-00001698, E.B.], by the Simons Foundation International [SFI-MPS-BH-00012593-02], and by the Italian Ministry of Foreign Affairs and International Cooperation Grant No.~PGR01167.
This work was carried out at the Advanced Research Computing at Hopkins (ARCH) core facility (\url{https://www.arch.jhu.edu/}), which is supported by the NSF Grant No.~OAC-1920103.
B.K. acknowledges support from the National Science Foundation Graduate Research Fellowship under Grant No. DGE-2236662. The work of J.K. and M.T. is supported in part by the DOE (HEP) Award No. DE-SC0013528.

\appendix
\section{Mass term in the wave equation: the role of vorticity in analog gravity}
\label{app vorticity}
\noindent
The results in the main text have been obtained considering the effective potential of a free massless field propagating on an ABH space-time. The purpose of this Appendix is to show how the TLN picture changes when including a component of vorticity in the background equation. As we will see, this results in an effective mass term in the perturbation equation. We will show that such an addition prevents the existence of the ladder structure. 

Instead of the canonical ABH in (3+1) dimensions considered in the main text, we will focus for simplicity on the ABH metric in (2+1) space-time dimensions, usually thought of as a draining bathtub, and studied in Ref.~\cite{DeLuca:2024nih}. As shown in Ref.~\cite{MalatoCorrea:2025iuc}, the presence of vorticity amounts to modifying the background velocity flow of the fluid according to
\begin{equation}
\vec v = \vec v_\text{\tiny irr} + \vec v_\text{\tiny rot}\,,
\end{equation}
where~$v_\text{\tiny irr}(v_\text{\tiny rot}$) denotes the irrotational (rotational) component. The background fluid vorticity thus reads
\begin{equation}
\vec \Omega = \vec \nabla \times \vec v_\text{\tiny rot} = \Omega \hat k\,,
\end{equation}
where~$\hat k$ is the unit vector along the axis of rotation, and~$\Omega = |\vec{\Omega}|$ is the magnitude of vorticity. 

Assuming solid body rotation ({\it i.e.}, constant~$\Omega$), it is possible to rewrite the equation of motion of the perturbation field~${\vec v}_\text{\tiny irr} = {\vec \nabla} \phi$ in the form~\cite{MalatoCorrea:2025iuc}
\begin{equation}
\label{vorticity}
\Box \phi = \frac{\Omega^2}{c_s^2} \phi\,,
\end{equation}
in terms of the effective metric 
\begin{equation}
{\rm d}s^2 = - \big(c_s^2 - |\vec v|^2\big) {\rm d}t^2 - 2 r \, v_\varphi\,  {\rm d} t {\rm d} \varphi + f(r)^{-1} {\rm d} r^2 + r^2 {\rm d} \varphi^2\,,
\end{equation}
where~$\varphi$ is the angular coordinate,~$v_r (v_\varphi)$ is the radial (angular) component of the background velocity flow, and~$f(r) = 1- v_r^2/c_s^2$.
Eq.~\eqref{vorticity} describes the propagation of a massive scalar field on a background geometry.

Assuming incompressibility, {\it i.e.},~$\vec \nabla \cdot \vec v=0$, one can rewrite the radial component as~$v_r = - c_s r_+/r$, by introducing the effective ABH horizon~$r_+$. Furthermore, assuming~$|v_\varphi/v_r| \ll 1$ allows us to bring the metric into a spherically symmetric form (by neglecting the~$t\varphi$ component), so that the equation of motion~\eqref{vorticity} becomes the equation of a massive perturbation, with effective mass proportional to~\cite{MalatoCorrea:2025iuc} 
\begin{equation}
\Omega = \frac{1}{r} \partial_r (r \, v_\varphi) = \frac{2 c_\varphi}{r_+}\,, \quad \text{with} \quad v_\varphi = c_\varphi \frac{r}{r_+}\,.
\end{equation}
Notice that the condition~$|v_\varphi/v_r| \ll 1$ translates into the requirement~$\frac{c_\varphi r^2}{c_s r_+^2} = \frac{\Omega^2 r^2}{4 c_s c_\varphi}  \ll 1$ on the size of the vorticity term~$c_\varphi$. 

In the small mass (vorticity) limit, one can approximate~$\Omega r \ll 1$, so that Eq.~\eqref{vorticity} admits a hypergeometric solution (in analogy to the near-zone metric of Schwarzschild black holes for the case of slowly varying time-dependent perturbations). By expanding the scalar field in multipole moments~$\phi = \sum_m R_m(r) {\rm e}^{{\rm i} m \varphi}$, the solution that is regular at the horizon thus reads~\cite{DeLuca:2024nih}
\begin{equation}
R_m (z) = C z^{|m_\Omega|}  {}_2F_1 (a,a-1, 2a -c; 1-z)\,,
\end{equation}
with~$z = r_+^2/r^2$,~$a = 1 + \frac{|m_\Omega|}{2}$,~$c = 1 + |m_\Omega|$ and the rescaled angular momentum coefficient
\begin{equation}
m_\Omega= \sqrt{m^2 - \frac{\Omega^2 r_+^2}{c_s^2}}\,.
\end{equation}
Following the derivation and results of Ref.~\cite{DeLuca:2024nih}, one can then extract the correction to the TLNs induced by vorticity to be (indicating by~$k_{m}^{(2+1) \, \text{\tiny ABH}}$ the result in the vorticity-free regime~$\Omega = 0$ of Ref.~\cite{DeLuca:2024nih})
\begin{widetext}
\begin{align}
k_{m_\Omega}^{(2+1) \, \text{\tiny ABH}} &= k_{m}^{(2+1) \, \text{\tiny ABH}} \nonumber \\
& + \frac{\Omega^2 r_+^2}{c_s^2}\begin{cases}
\frac{\Gamma (1-|m|) \Gamma \left(\frac{|m|}{2}\right)^2 \left[-\psi ^{(0)}\left(\frac{|m|}{2}\right)-\psi ^{(0)}\left(-\frac{|m|}{2}\right)+\psi ^{(0)}(-|m|)+\psi ^{(0)}(|m|)\right]}{8\Gamma \left(1-\frac{|m|}{2}\right)^2 \Gamma (|m|)} \quad \text{even} \,\, m\\
\frac{\pi  4^{-|m|} e^{-i \pi  |m|} \left[|m| \log (16)+2 |m| \psi ^{(0)}\left(\frac{|m|+1}{2}\right)-2 |m| \psi ^{(0)}\left(-\frac{|m|}{2}\right)+2\right] }{|m|^2 \Gamma \left(1-\frac{|m|}{2}\right) \Gamma \left(-\frac{|m|}{2}\right) \Gamma \left(\frac{|m|+1}{2}\right)^2} \log \left(\frac{r}{r_+}\right) \quad \text{odd} \,\, m
\end{cases} + \mathcal{O}\left(\frac{\Omega^4 r_+^4}{c_s^4} \right) \,.
\end{align}
\end{widetext}
This equation shows that the TLNs are always nonvanishing in the presence of an effective mass term in the equation of motion, thus invalidating the picture of a ladder structure to explain their vanishing value. 
We expect a similar conclusion to hold for canonical ABH in (3+1) dimensions.

\bibliography{draft}

@article{Fischer:2001jz,
    author = "Fischer, Uwe R. and Visser, Matt",
    title = "{Riemannian geometry of irrotational vortex acoustics}",
    eprint = "cond-mat/0110211",
    archivePrefix = "arXiv",
    doi = "10.1103/PhysRevLett.88.110201",
    journal = "Phys. Rev. Lett.",
    volume = "88",
    pages = "110201",
    year = "2002"
}

@article{Baak:2023zjf,
    author = "Baak, Sang-Shin and Datta, Satadal and Fischer, Uwe R.",
    title = "{Petrov classification of analogue spacetimes}",
    eprint = "2305.12771",
    archivePrefix = "arXiv",
    primaryClass = "gr-qc",
    doi = "10.1088/1361-6382/acf08e",
    journal = "Class. Quant. Grav.",
    volume = "40",
    number = "21",
    pages = "215001",
    year = "2023"
}

@article{Gupta:2001bg,
    author = "Gupta, Kumar S. and Sen, Siddhartha",
    title = "{Further evidence for the conformal structure of a Schwarzschild black hole in an algebraic approach}",
    eprint = "hep-th/0112041",
    archivePrefix = "arXiv",
    reportNumber = "SINP-TNP-01-28",
    doi = "10.1016/S0370-2693(01)01501-5",
    journal = "Phys. Lett. B",
    volume = "526",
    pages = "121--126",
    year = "2002"
}

@article{Bertini:2011ga,
    author = "Bertini, Stefano and Cacciatori, Sergio L. and Klemm, Dietmar",
    title = "{Conformal structure of the Schwarzschild black hole}",
    eprint = "1106.0999",
    archivePrefix = "arXiv",
    primaryClass = "hep-th",
    reportNumber = "IFUM-978-FT",
    doi = "10.1103/PhysRevD.85.064018",
    journal = "Phys. Rev. D",
    volume = "85",
    pages = "064018",
    year = "2012"
}

@article{MalatoCorrea:2025iuc,
    author = "Malato Corr{\^e}a, Mateus and Macedo, Caio F. B. and Panosso Macedo, Rodrigo and Oliveira, Leandro A.",
    title = "{Black hole spectral instabilities in the laboratory: Shallow water analog}",
    eprint = "2504.00107",
    archivePrefix = "arXiv",
    primaryClass = "gr-qc",
    doi = "10.1103/78ht-dn36",
    journal = "Phys. Rev. D",
    volume = "112",
    number = "2",
    pages = "024036",
    year = "2025"
}

@article{Berens:2025okm,
    author = "Berens, Roman and Hui, Lam and McLoughlin, Daniel and Penco, Riccardo and Staunton, John",
    title = "{Geometric Symmetries for the Vanishing of the Black Hole Tidal Love Numbers}",
    eprint = "2510.18952",
    archivePrefix = "arXiv",
    primaryClass = "hep-th",
    month = "10",
    year = "2025"
}

@article{Parra-Martinez:2025bcu,
    author = "Parra-Martinez, Julio and Podo, Alessandro",
    title = "{Naturalness of vanishing black-hole tides}",
    eprint = "2510.20694",
    archivePrefix = "arXiv",
    primaryClass = "hep-th",
    month = "10",
    year = "2025"
}

@article{Pereniguez:2025jxq,
    author = "Pere{\~n}iguez, David and Karnickis, Edgars",
    title = "{On the non-zero Love numbers of magnetic black holes}",
    eprint = "2509.12418",
    archivePrefix = "arXiv",
    primaryClass = "gr-qc",
    month = "9",
    year = "2025"
}

@article{DeLuca:2025bph,
    author = "De Luca, Valerio and Del Grosso, Loris and Iacovelli, Francesco and Maselli, Andrea and Berti, Emanuele",
    title = "{Systematic biases from ignoring environmental tidal effects in gravitational wave observations}",
    eprint = "2503.10746",
    archivePrefix = "arXiv",
    primaryClass = "gr-qc",
    doi = "10.1103/h4nh-nl5s",
    journal = "Phys. Rev. D",
    volume = "111",
    number = "12",
    pages = "124046",
    year = "2025"
}

@article{Chakravarti:2018vlt,
    author = "Chakravarti, Kabir and Chakraborty, Sumanta and Bose, Sukanta and SenGupta, Soumitra",
    title = "{Tidal Love numbers of black holes and neutron stars in the presence of higher dimensions: Implications of GW170817}",
    eprint = "1811.11364",
    archivePrefix = "arXiv",
    primaryClass = "gr-qc",
    doi = "10.1103/PhysRevD.99.024036",
    journal = "Phys. Rev. D",
    volume = "99",
    number = "2",
    pages = "024036",
    year = "2019"
}

@article{Chakravarti:2019aup,
    author = "Chakravarti, Kabir and Chakraborty, Sumanta and Phukon, Khun Sang and Bose, Sukanta and SenGupta, Soumitra",
    title = "{Constraining extra-spatial dimensions with observations of GW170817}",
    eprint = "1903.10159",
    archivePrefix = "arXiv",
    primaryClass = "gr-qc",
    reportNumber = "LIGO-P1900080",
    doi = "10.1088/1361-6382/ab8355",
    journal = "Class. Quant. Grav.",
    volume = "37",
    number = "10",
    pages = "105004",
    year = "2020"
}

@article{Caron-Huot:2025tlq,
    author = "Caron-Huot, Simon and Correia, Miguel and Isabella, Giulia and Solon, Mikhail",
    title = "{Gravitational Wave Scattering via the Born Series: Scalar Tidal Matching to O(G7) and Beyond}",
    eprint = "2503.13593",
    archivePrefix = "arXiv",
    primaryClass = "hep-th",
    doi = "10.1103/qd3c-nfz6",
    journal = "Phys. Rev. Lett.",
    volume = "135",
    number = "19",
    pages = "191601",
    year = "2025"
}

@article{Combaluzier-Szteinsznaider:2024sgb,
    author = "Combaluzier-Szteinsznaider, Oscar and Hui, Lam and Santoni, Luca and Solomon, Adam R. and Wong, Sam S. C.",
    title = "{Symmetries of vanishing nonlinear Love numbers of Schwarzschild black holes}",
    eprint = "2410.10952",
    archivePrefix = "arXiv",
    primaryClass = "gr-qc",
    doi = "10.1007/JHEP03(2025)124",
    journal = "JHEP",
    volume = "03",
    pages = "124",
    year = "2025"
}

@article{Chakraborty:2025wvs,
    author = "Chakraborty, Sumanta and De Luca, Valerio and Gualtieri, Leonardo and Pani, Paolo",
    title = "{Dynamical Love numbers of black holes: Theory and gravitational waveforms}",
    eprint = "2507.22994",
    archivePrefix = "arXiv",
    primaryClass = "gr-qc",
    doi = "10.1103/fr3y-s1sz",
    journal = "Phys. Rev. D",
    volume = "112",
    number = "10",
    pages = "104015",
    year = "2025"
}

@article{2006math.ph...2009O,
       author = {{Ovsienko}, Valentin},
        title = "{Large Coadjoint representation of Virasoro-type Lie algebras and differential operators on tensor-densities}",
      journal = {arXiv e-prints},
         year = 2006,
        month = feb,
          eid = {math-ph/0602009},
        pages = {math-ph/0602009},
          doi = {10.48550/arXiv.math-ph/0602009},
archivePrefix = {arXiv},
       eprint = {math-ph/0602009},
 primaryClass = {math-ph},
       adsurl = {https://ui.adsabs.harvard.edu/abs/2006math.ph...2009O}
}

@article{BeltranJimenez:2022hvs,
    author = "Beltr\'an Jim\'enez, Jose and Bettoni, Dario and Brax, Philippe",
    title = "{Polarisability and magnetisation of electrically K-mouflaged objects: the Born-Infeld ModMax case study}",
    eprint = "2211.16404",
    archivePrefix = "arXiv",
    primaryClass = "hep-th",
    reportNumber = "CERN-TH-2022-204",
    doi = "10.1007/JHEP02(2023)009",
    journal = "JHEP",
    volume = "02",
    pages = "009",
    year = "2023"
}

@article{BeltranJimenez:2024zmd,
    author = "Beltr\'an Jim\'enez, Jose and Bettoni, Dario and Brax, Philippe",
    title = "{Resilience of DBI screened objects and their ladder symmetries}",
    eprint = "2407.08627",
    archivePrefix = "arXiv",
    primaryClass = "hep-th",
    doi = "10.1007/JHEP10(2024)108",
    journal = "JHEP",
    volume = "10",
    pages = "108",
    year = "2024"
}

@article{Capuano:2024qhv,
    author = "Capuano, Lodovico and Santoni, Luca and Barausse, Enrico",
    title = "{Perturbations of the Vaidya metric in the frequency domain: Quasinormal modes and tidal response}",
    eprint = "2407.06009",
    archivePrefix = "arXiv",
    primaryClass = "gr-qc",
    doi = "10.1103/PhysRevD.110.084081",
    journal = "Phys. Rev. D",
    volume = "110",
    number = "8",
    pages = "084081",
    year = "2024"
}

@article{Vieira:2014rva,
    author = "Vieira, H. S. and Bezerra, V. B.",
    title = "{Acoustic black holes: massless scalar field analytic solutions and analogue Hawking radiation}",
    eprint = "1406.6884",
    archivePrefix = "arXiv",
    primaryClass = "gr-qc",
    doi = "10.1007/s10714-016-2082-x",
    journal = "Gen. Rel. Grav.",
    volume = "48",
    number = "7",
    pages = "88",
    year = "2016",
    note = "[Erratum: Gen.Rel.Grav. 51, 51 (2019)]"
}

@article{Carlip:1998wz,
    author = "Carlip, Steven",
    title = "{Black hole entropy from conformal field theory in any dimension}",
    eprint = "hep-th/9812013",
    archivePrefix = "arXiv",
    reportNumber = "UCD-98-18",
    doi = "10.1103/PhysRevLett.82.2828",
    journal = "Phys. Rev. Lett.",
    volume = "82",
    pages = "2828--2831",
    year = "1999"
}

@article{Gounis:2024hcm,
    author = "Gounis, L. -R. and Kehagias, A. and Riotto, A.",
    title = "{The vanishing of the non-linear static love number of Kerr black holes and the role of symmetries}",
    eprint = "2412.08249",
    archivePrefix = "arXiv",
    primaryClass = "gr-qc",
    doi = "10.1088/1475-7516/2025/03/002",
    journal = "JCAP",
    volume = "03",
    pages = "002",
    year = "2025"
}

@article{Bhatt:2024yyz,
    author = "Bhatt, Rajendra Prasad and Chakraborty, Sumanta and Bose, Sukanta",
    title = "{Rotating black holes experience dynamical tides}",
    eprint = "2406.09543",
    archivePrefix = "arXiv",
    primaryClass = "gr-qc",
    reportNumber = "LIGO-P2400258",
    doi = "10.1103/PhysRevD.111.L041504",
    journal = "Phys. Rev. D",
    volume = "111",
    number = "4",
    pages = "L041504",
    year = "2025"
}

@article{Berens:2025jfs,
    author = "Berens, Roman and Hui, Lam and McLoughlin, Daniel and Solomon, Adam R. and Staunton, John",
    title = "{Ladder Symmetries of Higher Dimensional Black Holes}",
    eprint = "2510.26748",
    archivePrefix = "arXiv",
    primaryClass = "hep-th",
    month = "10",
    year = "2025"
}

@article{Kobayashi:2025vgl,
    author = "Kobayashi, Hajime and Mukohyama, Shinji and Oshita, Naritaka and Takahashi, Kazufumi and Yingcharoenrat, Vicharit",
    title = "{Dynamical Tidal Response of Non-rotating Black Holes: Connecting the MST Formalism and Worldline EFT}",
    eprint = "2511.12580",
    archivePrefix = "arXiv",
    primaryClass = "gr-qc",
    reportNumber = "YITP-25-176, RESCEU-25/25, IPMU25-0052, RIKEN-iTHEMS-Report-25",
    month = "11",
    year = "2025"
}

@article{Rodriguez:2025ala,
    author = "Rodriguez, Maria J. and Temoche, Luis Fernando",
    title = "{Clines and the Analytic Structure of Black Hole Perturbations}",
    eprint = "2511.08695",
    archivePrefix = "arXiv",
    primaryClass = "gr-qc",
    month = "11",
    year = "2025"
}

@article{Gounis:2025tmt,
    author = "Gounis, L. R. and Kehagias, A. and Panagopoulos, G. and Riotto, A.",
    title = "{Non-vanishing non-linear Static Love Number of a Class of Extremal Reissner-Nordstrom Black Holes}",
    eprint = "2512.02506",
    archivePrefix = "arXiv",
    primaryClass = "gr-qc",
    month = "12",
    year = "2025"
}

@article{Combaluzier--Szteinsznaider:2025eoc,
    author = "Combaluzier--Szteinsznaider, Oscar and Glazer, Daniel and Joyce, Austin and Rodriguez, Maria J. and Santoni, Luca",
    title = "{Dynamical Tidal Response of Schwarzschild Black Holes}",
    eprint = "2511.02372",
    archivePrefix = "arXiv",
    primaryClass = "gr-qc",
    month = "11",
    year = "2025"
}

@article{Essin:2006sic,
    author = "Essin, Andrew M. and Griffiths, David J.",
    title = "{Quantum mechanics of the 1{\ensuremath{/}}x2 potential}",
    doi = "10.1119/1.2165248",
    journal = "Am. J. Phys.",
    volume = "74",
    number = "2",
    pages = "109",
    year = "2006"
}

@article{Coviello:2025pla,
    author = "Coviello, Chiara and Vellucci, Vania and Lehner, Luis",
    title = "{Tidal response of regular black holes}",
    eprint = "2503.04287",
    archivePrefix = "arXiv",
    primaryClass = "gr-qc",
    doi = "10.1103/PhysRevD.111.104073",
    journal = "Phys. Rev. D",
    volume = "111",
    number = "10",
    pages = "104073",
    year = "2025"
}

@article{Avitan:2023txy,
    author = "Avitan, Shani and Brustein, Ram and Sherf, Yotam",
    title = "{Discovering Love numbers through resonance excitation during extreme mass ratio inspirals}",
    eprint = "2306.00173",
    archivePrefix = "arXiv",
    primaryClass = "gr-qc",
    doi = "10.1088/1361-6382/ad563a",
    journal = "Class. Quant. Grav.",
    volume = "41",
    number = "14",
    pages = "145009",
    year = "2024"
}

@article{Donnay:2016ejv,
    author = "Donnay, Laura and Giribet, Gaston and Gonz\'alez, Hern\'an A. and Pino, Miguel",
    title = "{Extended Symmetries at the Black Hole Horizon}",
    eprint = "1607.05703",
    archivePrefix = "arXiv",
    primaryClass = "hep-th",
    doi = "10.1007/JHEP09(2016)100",
    journal = "JHEP",
    volume = "09",
    pages = "100",
    year = "2016"
}

@article{Akhmedov:2017ftb,
    author = "Akhmedov, Emil T. and Godazgar, Mahdi",
    title = "{Symmetries at the black hole horizon}",
    eprint = "1707.05517",
    archivePrefix = "arXiv",
    primaryClass = "hep-th",
    doi = "10.1103/PhysRevD.96.104025",
    journal = "Phys. Rev. D",
    volume = "96",
    number = "10",
    pages = "104025",
    year = "2017"
}

@article{Carlip:2011vr,
    author = "Carlip, Steven",
    title = "{Effective Conformal Descriptions of Black Hole Entropy}",
    eprint = "1107.2678",
    archivePrefix = "arXiv",
    primaryClass = "gr-qc",
    doi = "10.3390/e13071355",
    journal = "Entropy",
    volume = "13",
    pages = "1355--1379",
    year = "2011"
}

@article{Cadoni:2005ej,
    author = "Cadoni, Mariano",
    title = "{Statistical entropy of the Schwarzschild black hole}",
    eprint = "hep-th/0511103",
    archivePrefix = "arXiv",
    doi = "10.1142/S0217732306021165",
    journal = "Mod. Phys. Lett. A",
    volume = "21",
    pages = "1879--1888",
    year = "2006"
}

@article{Hotta:2000gx,
    author = "Hotta, M. and Sasaki, K. and Sasaki, T.",
    title = "{Diffeomorphism on horizon as an asymptotic isometry of Schwarzschild black hole}",
    eprint = "gr-qc/0011043",
    archivePrefix = "arXiv",
    reportNumber = "TU-606",
    doi = "10.1088/0264-9381/18/10/301",
    journal = "Class. Quant. Grav.",
    volume = "18",
    pages = "1823--1834",
    year = "2001"
}

@article{DeLuca:2024nih,
    author = "De Luca, Valerio and Khek, Brandon and Khoury, Justin and Trodden, Mark",
    title = "{Tidal Love numbers of analog black holes}",
    eprint = "2412.08728",
    archivePrefix = "arXiv",
    primaryClass = "gr-qc",
    doi = "10.1103/PhysRevD.111.044069",
    journal = "Phys. Rev. D",
    volume = "111",
    number = "4",
    pages = "044069",
    year = "2025"
}

@article{Cardoso:2017egd,
    author = "Cardoso, Vitor and Houri, Tsuyoshi and Kimura, Masashi",
    title = "{General first-order mass ladder operators for Klein\textendash{}Gordon fields}",
    eprint = "1707.08534",
    archivePrefix = "arXiv",
    primaryClass = "hep-th",
    reportNumber = "KUNS-2692",
    doi = "10.1088/1361-6382/aa9a04",
    journal = "Class. Quant. Grav.",
    volume = "35",
    number = "1",
    pages = "015011",
    year = "2018"
}

@article{Cardoso:2017qmj,
    author = "Cardoso, Vitor and Houri, Tsuyoshi and Kimura, Masashi",
    title = "{Mass Ladder Operators from Spacetime Conformal Symmetry}",
    eprint = "1706.07339",
    archivePrefix = "arXiv",
    primaryClass = "hep-th",
    reportNumber = "KOBE-COSMO-17-07, KUNS-2683",
    doi = "10.1103/PhysRevD.96.024044",
    journal = "Phys. Rev. D",
    volume = "96",
    number = "2",
    pages = "024044",
    year = "2017"
}

@article{Vieira:2021xqw,
    author = "Vieira, H. S. and Kokkotas, Kostas D.",
    title = "{Quasibound states of Schwarzschild acoustic black holes}",
    eprint = "2104.03938",
    archivePrefix = "arXiv",
    primaryClass = "gr-qc",
    doi = "10.1103/PhysRevD.104.024035",
    journal = "Phys. Rev. D",
    volume = "104",
    number = "2",
    pages = "024035",
    year = "2021"
}

@article{Basak:2002aw,
    author = "Basak, Soumen and Majumdar, Parthasarathi",
    title = "{`Superresonance' from a rotating acoustic black hole}",
    eprint = "gr-qc/0203059",
    archivePrefix = "arXiv",
    doi = "10.1088/0264-9381/20/18/304",
    journal = "Class. Quant. Grav.",
    volume = "20",
    pages = "3907--3914",
    year = "2003"
}

@article{Vieira:2021ozg,
    author = "Vieira, H. S. and Destounis, Kyriakos and Kokkotas, Kostas D.",
    title = "{Slowly-rotating curved acoustic black holes: Quasinormal modes, Hawking-Unruh radiation, and quasibound states}",
    eprint = "2112.08711",
    archivePrefix = "arXiv",
    primaryClass = "gr-qc",
    doi = "10.1103/PhysRevD.105.045015",
    journal = "Phys. Rev. D",
    volume = "105",
    number = "4",
    pages = "045015",
    year = "2022"
}

@article{Unruh:1980cg,
    author = "Unruh, W. G.",
    title = "{Experimental black hole evaporation}",
    doi = "10.1103/PhysRevLett.46.1351",
    journal = "Phys. Rev. Lett.",
    volume = "46",
    pages = "1351--1353",
    year = "1981"
}

@proceedings{Deruelle:1984hq,
      editor         = "Deruelle, N. and Piran, T.",
      title          = "{Gravitational Radiation. Proceedings, Summer School,
                        NATO Advanced Study Institute, Les Houches, France, June 2-21, 1982}",
      journal        = "Amsterdam, Netherlands: North-holland ( 1983) 510p",
      year           = "1984",
      SLACcitation   = "%%CITATION = INSPIRE-203412;%%"
}

@article{Brito:2023pyl,
    author = "Brito, Richard and Shah, Shreya",
    title = "{Extreme mass-ratio inspirals into black holes surrounded by scalar clouds}",
    eprint = "2307.16093",
    archivePrefix = "arXiv",
    primaryClass = "gr-qc",
    doi = "10.1103/PhysRevD.108.084019",
    journal = "Phys. Rev. D",
    volume = "108",
    number = "8",
    pages = "084019",
    year = "2023"
}

@article{DeLuca:2024ufn,
    author = "De Luca, Valerio and Garoffolo, Alice and Khoury, Justin and Trodden, Mark",
    title = "{Tidal Love numbers and Green\textquoteright{}s functions in black hole spacetimes}",
    eprint = "2407.07156",
    archivePrefix = "arXiv",
    primaryClass = "gr-qc",
    doi = "10.1103/PhysRevD.110.064081",
    journal = "Phys. Rev. D",
    volume = "110",
    number = "6",
    pages = "064081",
    year = "2024"
}

@article{Camblong:2003mz,
    author = "Camblong, Horacio E. and Ordonez, Carlos R.",
    title = "{Anomaly in conformal quantum mechanics: From molecular physics to black holes}",
    eprint = "hep-th/0303166",
    archivePrefix = "arXiv",
    doi = "10.1103/PhysRevD.68.125013",
    journal = "Phys. Rev. D",
    volume = "68",
    pages = "125013",
    year = "2003"
}

@article{Camblong:2003mb,
    author = "Camblong, Horacio E. and Ordonez, Carlos R.",
    title = "{Renormalization in conformal quantum mechanics}",
    eprint = "hep-th/0305035",
    archivePrefix = "arXiv",
    doi = "10.1016/j.physleta.2005.06.110",
    journal = "Phys. Lett. A",
    volume = "345",
    pages = "22--30",
    year = "2005"
}

@article{Govindarajan:2000ag,
    author = "Govindarajan, T. R. and Suneeta, V. and Vaidya, S.",
    title = "{Horizon states for AdS black holes}",
    eprint = "hep-th/0002036",
    archivePrefix = "arXiv",
    reportNumber = "IMSC-2000-02-03, TIFR-TH-00-08",
    doi = "10.1016/S0550-3213(00)00336-9",
    journal = "Nucl. Phys. B",
    volume = "583",
    pages = "291--303",
    year = "2000"
}

@article{Birmingham:2001qa,
    author = "Birmingham, Danny and Gupta, Kumar S. and Sen, Siddhartha",
    title = "{Near horizon conformal structure of black holes}",
    eprint = "hep-th/0102051",
    archivePrefix = "arXiv",
    reportNumber = "SINP-TNP-01-02",
    doi = "10.1016/S0370-2693(01)00354-9",
    journal = "Phys. Lett. B",
    volume = "505",
    pages = "191--196",
    year = "2001"
}

@book{osti_4271130,
  author       = {Wybourne, B G},
  title        = {Classical groups for physicists},
  url          = {https://www.osti.gov/biblio/4271130},
  place        = {Country unknown/Code not available},
  publisher    = {John Wiley and Sons, Inc., New York},
  year         = {1974},
  month        = {01}}

@article{Camblong:2004ye,
    author = "Camblong, Horacio E. and Ordonez, Carlos R.",
    title = "{Black hole thermodynamics from near-horizon conformal quantum mechanics}",
    eprint = "hep-th/0411008",
    archivePrefix = "arXiv",
    doi = "10.1103/PhysRevD.71.104029",
    journal = "Phys. Rev. D",
    volume = "71",
    pages = "104029",
    year = "2005"
}

@article{Horvathy:2009kz,
    author = "Horvathy, P. A. and Zhang, P. -M.",
    title = "{Non-relativistic conformal symmetries in fluid mechanics}",
    eprint = "0906.3594",
    archivePrefix = "arXiv",
    primaryClass = "physics.flu-dyn",
    doi = "10.1140/epjc/s10052-009-1221-x",
    journal = "Eur. Phys. J. C",
    volume = "65",
    pages = "607--614",
    year = "2010"
}

@article{deAlfaro:1976vlx,
    author = "de Alfaro, Vittorio and Fubini, S. and Furlan, G.",
    title = "{Conformal Invariance in Quantum Mechanics}",
    reportNumber = "CERN-TH-2115",
    doi = "10.1007/BF02785666",
    journal = "Nuovo Cim. A",
    volume = "34",
    pages = "569",
    year = "1976"
}

@article{DeLuca:2024uju,
    author = "De Luca, Valerio and Franciolini, Gabriele and Riotto, Antonio",
    title = "{Flea on the elephant: Tidal Love numbers in subsolar primordial black hole searches}",
    eprint = "2408.14207",
    archivePrefix = "arXiv",
    primaryClass = "gr-qc",
    doi = "10.1103/PhysRevD.110.104041",
    journal = "Phys. Rev. D",
    volume = "110",
    number = "10",
    pages = "104041",
    year = "2024"
}

@article{Katagiri:2022vyz,
    author = "Katagiri, Takuya and Kimura, Masashi and Nakano, Hiroyuki and Omukai, Kazuyuki",
    title = "{Vanishing Love numbers of black holes in general relativity: From spacetime conformal symmetry of a two-dimensional reduced geometry}",
    eprint = "2209.10469",
    archivePrefix = "arXiv",
    primaryClass = "gr-qc",
    doi = "10.1103/PhysRevD.107.124030",
    journal = "Phys. Rev. D",
    volume = "107",
    number = "12",
    pages = "124030",
    year = "2023"
}

@article{Ma:2024few,
    author = "Ma, Liang and Wu, Ze-Hua and Pang, Yi and Lu, H.",
    title = "{Charging the Love numbers: Charged scalar response coefficients of Kerr-Newman black holes}",
    eprint = "2408.10352",
    archivePrefix = "arXiv",
    primaryClass = "gr-qc",
    doi = "10.1103/PhysRevD.111.044003",
    journal = "Phys. Rev. D",
    volume = "111",
    number = "4",
    pages = "044003",
    year = "2025"
}

@article{Colpi:2024xhw,
    author = "Colpi, Monica and others",
    title = "{LISA Definition Study Report}",
    eprint = "2402.07571",
    archivePrefix = "arXiv",
    primaryClass = "astro-ph.CO",
    month = "2",
    year = "2024"
}

@article{Reitze:2019iox,
	title        = {{Cosmic Explorer: The U.S. Contribution to Gravitational-Wave Astronomy beyond LIGO}},
	author       = {Reitze, David and others},
	year         = 2019,
	journal      = {Bull. Am. Astron. Soc.},
	volume       = 51,
	number       = 7,
	pages        = {035},
	eprint       = {1907.04833},
	archiveprefix = {arXiv},
	primaryclass = {astro-ph.IM},
	reportnumber = {LIGO-P1900316}
}

@article{Katagiri:2023yzm,
    author = "Katagiri, Takuya and Nakano, Hiroyuki and Omukai, Kazuyuki",
    title = "{Stability of relativistic tidal response against small potential modification}",
    eprint = "2304.04551",
    archivePrefix = "arXiv",
    primaryClass = "gr-qc",
    doi = "10.1103/PhysRevD.108.084049",
    journal = "Phys. Rev. D",
    volume = "108",
    number = "8",
    pages = "084049",
    year = "2023"
}

@article{Barura:2024uog,
    author = "Barura, Chams Gharib Ali and Kobayashi, Hajime and Mukohyama, Shinji and Oshita, Naritaka and Takahashi, Kazufumi and Yingcharoenrat, Vicharit",
    title = "{Tidal Love numbers from EFT of black hole perturbations with timelike scalar profile}",
    eprint = "2405.10813",
    archivePrefix = "arXiv",
    primaryClass = "gr-qc",
    reportNumber = "YITP-24-62, IPMU24-0018, RIKEN-iTHEMS-Report-24",
    doi = "10.1088/1475-7516/2024/09/001",
    journal = "JCAP",
    volume = "09",
    pages = "001",
    year = "2024"
}

@article{DeLuca:2021ite,
	title        = {{Tidal deformability of dressed black holes and tests of ultralight bosons in extended mass ranges}},
	author       = {De Luca, Valerio and Pani, Paolo},
	year         = 2021,
	journal      = {JCAP},
	volume       = {08},
	pages        = {032},
	doi          = {10.1088/1475-7516/2021/08/032},
	eprint       = {2106.14428},
	archiveprefix = {arXiv},
	primaryclass = {gr-qc}
}

@article{Lenzi:2021wpc,
    author = "Lenzi, Michele and Sopuerta, Carlos F.",
    title = "{Master functions and equations for perturbations of vacuum spherically symmetric spacetimes}",
    eprint = "2108.08668",
    archivePrefix = "arXiv",
    primaryClass = "gr-qc",
    doi = "10.1103/PhysRevD.104.084053",
    journal = "Phys. Rev. D",
    volume = "104",
    number = "8",
    pages = "084053",
    year = "2021"
}

@article{Glampedakis:2017rar,
    author = "Glampedakis, Kostas and Johnson, Aaron D. and Kennefick, Daniel",
    title = "{Darboux transformation in black hole perturbation theory}",
    eprint = "1702.06459",
    archivePrefix = "arXiv",
    primaryClass = "gr-qc",
    doi = "10.1103/PhysRevD.96.024036",
    journal = "Phys. Rev. D",
    volume = "96",
    number = "2",
    pages = "024036",
    year = "2017"
}

@article{Cooper:1994eh,
    author = "Cooper, Fred and Khare, Avinash and Sukhatme, Uday",
    title = "{Supersymmetry and quantum mechanics}",
    eprint = "hep-th/9405029",
    archivePrefix = "arXiv",
    reportNumber = "LA-UR-94-569",
    doi = "10.1016/0370-1573(94)00080-M",
    journal = "Phys. Rept.",
    volume = "251",
    pages = "267--385",
    year = "1995"
}

@article{1999physics...8003D,
       author = {{Darboux}, G.},
        title = "{On a proposition relative to linear equations}",
      journal = {arXiv e-prints},
         year = 1999,
        month = aug,
          eid = {physics/9908003},
        pages = {physics/9908003},
          doi = {10.48550/arXiv.physics/9908003},
archivePrefix = {arXiv},
       eprint = {physics/9908003},
 primaryClass = {physics.hist-ph},
       adsurl = {https://ui.adsabs.harvard.edu/abs/1999physics...8003D}
}

@article{Chandrasekhar:1979iz,
    author = "Chandrasekhar, Subrahmanyan",
    editor = "Wali, K. C.",
    title = "{ON THE EQUATIONS GOVERNING THE PERTURBATIONS OF THE REISSNER-NORDSTROM BLACK HOLE}",
    doi = "10.1098/rspa.1979.0028",
    journal = "Proc. Roy. Soc. Lond. A",
    volume = "365",
    pages = "453--465",
    year = "1979"
}

@article{Chandrasekhar:1975zza,
    author = "Chandrasekhar, S. and Detweiler, Steven L.",
    title = "{The quasi-normal modes of the Schwarzschild black hole}",
    doi = "10.1098/rspa.1975.0112",
    journal = "Proc. Roy. Soc. Lond. A",
    volume = "344",
    pages = "441--452",
    year = "1975"
}

@article{Riva:2023rcm,
    author = "Riva, Massimiliano Maria and Santoni, Luca and Savi\'c, Nikola and Vernizzi, Filippo",
    title = "{Vanishing of nonlinear tidal Love numbers of Schwarzschild black holes}",
    eprint = "2312.05065",
    archivePrefix = "arXiv",
    primaryClass = "gr-qc",
    doi = "10.1016/j.physletb.2024.138710",
    journal = "Phys. Lett. B",
    volume = "854",
    pages = "138710",
    year = "2024"
}

@article{Chakraborty:2023zed,
    author = "Chakraborty, Sumanta and Maggio, Elisa and Silvestrini, Michela and Pani, Paolo",
    title = "{Dynamical tidal Love numbers of Kerr-like compact objects}",
    eprint = "2310.06023",
    archivePrefix = "arXiv",
    primaryClass = "gr-qc",
    doi = "10.1103/PhysRevD.110.084042",
    journal = "Phys. Rev. D",
    volume = "110",
    number = "8",
    pages = "084042",
    year = "2024"
}

@mastersthesis{Charalambous:2024gpf,
    author = "Charalambous, Panagiotis",
    title = "{Magic zeroes in the black hole response problem and a Love symmetry resolution}",
    eprint = "2404.17030",
    archivePrefix = "arXiv",
    primaryClass = "hep-th",
    type = "Other thesis",
    month = "4",
    year = "2024"
}

@article{Kehagias:2025tqi,
    author = "Kehagias, Alex and Riotto, Antonio",
    title = "{AdS perspective on the nonlinear tails in black hole ringdowns}",
    eprint = "2506.14475",
    archivePrefix = "arXiv",
    primaryClass = "gr-qc",
    doi = "10.1103/yk6g-sdbw",
    journal = "Phys. Rev. D",
    volume = "112",
    number = "8",
    pages = "084068",
    year = "2025"
}

@article{Nair:2022xfm,
    author = "Nair, Sreejith and Chakraborty, Sumanta and Sarkar, Sudipta",
    title = "{Dynamical Love numbers for area quantized black holes}",
    eprint = "2208.06235",
    archivePrefix = "arXiv",
    primaryClass = "gr-qc",
    doi = "10.1103/PhysRevD.107.124041",
    journal = "Phys. Rev. D",
    volume = "107",
    number = "12",
    pages = "124041",
    year = "2023"
}

@article{Cannizzaro:2024fpz,
    author = "Cannizzaro, Enrico and De Luca, Valerio and Pani, Paolo",
    title = "{Tidal deformability of black holes surrounded by thin accretion disks}",
    eprint = "2408.14208",
    archivePrefix = "arXiv",
    primaryClass = "astro-ph.HE",
    doi = "10.1103/PhysRevD.110.123004",
    journal = "Phys. Rev. D",
    volume = "110",
    number = "12",
    pages = "123004",
    year = "2024"
}

@article{Anderson:1991kx,
    author = "Anderson, A. and Price, R. H.",
    title = "{Intertwining of the equations of black hole perturbations}",
    doi = "10.1103/PhysRevD.43.3147",
    journal = "Phys. Rev. D",
    volume = "43",
    pages = "3147--3154",
    year = "1991"
}

@phdthesis{Katsimpouri:2015nqc,
    author = "Katsimpouri, Despoina",
    title = "{Integrability in two-dimensional gravity}",
    school = "Humboldt U., Berlin",
    year = "2015"
}

@article{Lu:2007zv,
    author = "Lu, H. and Perry, M. J. and Pope, C. N.",
    title = "{Infinite-dimensional symmetries of two-dimensional coset models}",
    eprint = "0711.0400",
    archivePrefix = "arXiv",
    primaryClass = "hep-th",
    reportNumber = "DAMTP-2007-107, MIFP-07-28",
    month = "11",
    year = "2007"
}

@article{Lu:2007jc,
    author = "Lu, H. and Perry, M. J. and Pope, C. N.",
    title = "{Infinite-dimensional symmetries of two-dimensional coset models coupled to gravity}",
    eprint = "0712.0615",
    archivePrefix = "arXiv",
    primaryClass = "hep-th",
    reportNumber = "DAMTP-2007-115, MIFP-07-31",
    doi = "10.1016/j.nuclphysb.2008.07.035",
    journal = "Nucl. Phys. B",
    volume = "806",
    pages = "656--683",
    year = "2009"
}

@article{Maison:2000fj,
    author = "Maison, D.",
    editor = "Schmidt, Bernd G.",
    title = "{Duality and hidden symmetries in gravitational theories}",
    journal = "Lect. Notes Phys.",
    volume = "540",
    pages = "273--323",
    year = "2000"
}

@article{Breitenlohner:1986um,
    author = "Breitenlohner, Peter and Maison, Dieter",
    title = "{On the Geroch Group}",
    reportNumber = "MPI-PAE/PTh-70/86",
    journal = "Ann. Inst. H. Poincare Phys. Theor.",
    volume = "46",
    pages = "215",
    year = "1987"
}

@article{Lenzi:2021njy,
    author = "Lenzi, Michele and Sopuerta, Carlos F.",
    title = "{Darboux covariance: A hidden symmetry of perturbed Schwarzschild black holes}",
    eprint = "2109.00503",
    archivePrefix = "arXiv",
    primaryClass = "gr-qc",
    doi = "10.1103/PhysRevD.104.124068",
    journal = "Phys. Rev. D",
    volume = "104",
    number = "12",
    pages = "124068",
    year = "2021"
}

@article{Geroch:1970nt,
    author = "Geroch, Robert P.",
    title = "{A Method for generating solutions of Einstein's equations}",
    doi = "10.1063/1.1665681",
    journal = "J. Math. Phys.",
    volume = "12",
    pages = "918--924",
    year = "1971"
}

@article{Creci:2021rkz,
    author = "Creci, Gast\'on and Hinderer, Tanja and Steinhoff, Jan",
    title = "{Tidal response from scattering and the role of analytic continuation}",
    eprint = "2108.03385",
    archivePrefix = "arXiv",
    primaryClass = "gr-qc",
    doi = "10.1103/PhysRevD.104.124061",
    journal = "Phys. Rev. D",
    volume = "104",
    number = "12",
    pages = "124061",
    year = "2021",
    note = "[Erratum: Phys.Rev.D 105, 109902 (2022)]"
}

@article{Perry:2023wmm,
    author = "Perry, Malcolm and Rodriguez, Maria J.",
    title = "{Dynamical Love Numbers for Kerr Black Holes}",
    eprint = "2310.03660",
    archivePrefix = "arXiv",
    primaryClass = "gr-qc",
    month = "10",
    year = "2023"
}

@article{Hui:2021vcv,
    author = "Hui, Lam and Joyce, Austin and Penco, Riccardo and Santoni, Luca and Solomon, Adam R.",
    title = "{Ladder symmetries of black holes. Implications for love numbers and no-hair theorems}",
    eprint = "2105.01069",
    archivePrefix = "arXiv",
    primaryClass = "hep-th",
    doi = "10.1088/1475-7516/2022/01/032",
    journal = "JCAP",
    volume = "01",
    number = "01",
    pages = "032",
    year = "2022"
}

@article{Jackiw:1972cb,
    author = "Jackiw, R.",
    title = "{Introducing scale symmetry}",
    doi = "10.1063/1.3070673",
    journal = "Phys. Today",
    volume = "25N1",
    pages = "23--27",
    year = "1972"
}

@article{BenAchour:2022uqo,
    author = "Ben Achour, Jibril and Livine, Etera R. and Mukohyama, Shinji and Uzan, Jean-Philippe",
    title = "{Hidden symmetry of the static response of black holes: applications to Love numbers}",
    eprint = "2202.12828",
    archivePrefix = "arXiv",
    primaryClass = "gr-qc",
    reportNumber = "YITP-22-17, IPMU22-0003",
    doi = "10.1007/JHEP07(2022)112",
    journal = "JHEP",
    volume = "07",
    pages = "112",
    year = "2022"
}

@article{Baumann:2018vus,
    author = "Baumann, Daniel and Chia, Horng Sheng and Porto, Rafael A.",
    title = "{Probing Ultralight Bosons with Binary Black Holes}",
    eprint = "1804.03208",
    archivePrefix = "arXiv",
    primaryClass = "gr-qc",
    reportNumber = "DESY-18-060, DESY 18-060",
    doi = "10.1103/PhysRevD.99.044001",
    journal = "Phys. Rev. D",
    volume = "99",
    number = "4",
    pages = "044001",
    year = "2019"
}

@article{Cardoso:2019upw,
    author = "Cardoso, Vitor and Duque, Francisco",
    title = "{Environmental effects in gravitational-wave physics: Tidal deformability of black holes immersed in matter}",
    eprint = "1912.07616",
    archivePrefix = "arXiv",
    primaryClass = "gr-qc",
    doi = "10.1103/PhysRevD.101.064028",
    journal = "Phys. Rev. D",
    volume = "101",
    number = "6",
    pages = "064028",
    year = "2020"
}

@article{DeLuca:2022xlz,
    author = "De Luca, Valerio and Maselli, Andrea and Pani, Paolo",
    title = "{Modeling frequency-dependent tidal deformability for environmental black hole mergers}",
    eprint = "2212.03343",
    archivePrefix = "arXiv",
    primaryClass = "gr-qc",
    doi = "10.1103/PhysRevD.107.044058",
    journal = "Phys. Rev. D",
    volume = "107",
    number = "4",
    pages = "044058",
    year = "2023"
}

@article{Berens:2022ebl,
    author = "Berens, Roman and Hui, Lam and Sun, Zimo",
    title = "{Ladder symmetries of black holes and de Sitter space: love numbers and quasinormal modes}",
    eprint = "2212.09367",
    archivePrefix = "arXiv",
    primaryClass = "hep-th",
    doi = "10.1088/1475-7516/2023/06/056",
    journal = "JCAP",
    volume = "06",
    pages = "056",
    year = "2023"
}

@article{Rodriguez:2023xjd,
    author = "Rodriguez, Maria J. and Santoni, Luca and Solomon, Adam R. and Temoche, Luis Fernando",
    title = "{Love numbers for rotating black holes in higher dimensions}",
    eprint = "2304.03743",
    archivePrefix = "arXiv",
    primaryClass = "hep-th",
    doi = "10.1103/PhysRevD.108.084011",
    journal = "Phys. Rev. D",
    volume = "108",
    number = "8",
    pages = "084011",
    year = "2023"
}

@article{Ivanov:2024sds,
    author = "Ivanov, Mikhail M. and Li, Yue-Zhou and Parra-Martinez, Julio and Zhou, Zihan",
    title = "{Gravitational Raman Scattering in Effective Field Theory: A Scalar Tidal Matching at O(G3)}",
    eprint = "2401.08752",
    archivePrefix = "arXiv",
    primaryClass = "hep-th",
    reportNumber = "MIT-CTP/5664",
    doi = "10.1103/PhysRevLett.132.131401",
    journal = "Phys. Rev. Lett.",
    volume = "132",
    number = "13",
    pages = "131401",
    year = "2024"
}

@article{DeLuca:2022tkm,
    author = "De Luca, Valerio and Khoury, Justin and Wong, Sam S. C.",
    title = "{Implications of the weak gravity conjecture for tidal Love numbers of black holes}",
    eprint = "2211.14325",
    archivePrefix = "arXiv",
    primaryClass = "hep-th",
    doi = "10.1103/PhysRevD.108.044066",
    journal = "Phys. Rev. D",
    volume = "108",
    number = "4",
    pages = "044066",
    year = "2023"
}

@article{Pereniguez:2021xcj,
    author = "Pere\~niguez, David and Cardoso, Vitor",
    title = "{Love numbers and magnetic susceptibility of charged black holes}",
    eprint = "2112.08400",
    archivePrefix = "arXiv",
    primaryClass = "gr-qc",
    doi = "10.1103/PhysRevD.105.044026",
    journal = "Phys. Rev. D",
    volume = "105",
    number = "4",
    pages = "044026",
    year = "2022"
}

@article{Bonelli:2021uvf,
    author = "Bonelli, Giulio and Iossa, Cristoforo and Lichtig, Daniel Panea and Tanzini, Alessandro",
    title = "{Exact solution of Kerr black hole perturbations via CFT2 and instanton counting: Greybody factor, quasinormal modes, and Love numbers}",
    eprint = "2105.04483",
    archivePrefix = "arXiv",
    primaryClass = "hep-th",
    doi = "10.1103/PhysRevD.105.044047",
    journal = "Phys. Rev. D",
    volume = "105",
    number = "4",
    pages = "044047",
    year = "2022"
}

@article{Ivanov:2022qqt,
    author = "Ivanov, Mikhail M. and Zhou, Zihan",
    title = "{Vanishing of Black Hole Tidal Love Numbers from Scattering Amplitudes}",
    eprint = "2209.14324",
    archivePrefix = "arXiv",
    primaryClass = "hep-th",
    doi = "10.1103/PhysRevLett.130.091403",
    journal = "Phys. Rev. Lett.",
    volume = "130",
    number = "9",
    pages = "091403",
    year = "2023"
}

@article{Charalambous:2022rre,
    author = "Charalambous, Panagiotis and Dubovsky, Sergei and Ivanov, Mikhail M.",
    title = "{Love symmetry}",
    eprint = "2209.02091",
    archivePrefix = "arXiv",
    primaryClass = "hep-th",
    doi = "10.1007/JHEP10(2022)175",
    journal = "JHEP",
    volume = "10",
    pages = "175",
    year = "2022"
}

@article{Damour:2009va,
	author = "Damour, Thibault and Lecian, Orchidea Maria",
	title = "{On the gravitational polarizability of black holes}",
	eprint = "0906.3003",
	archivePrefix = "arXiv",
	primaryClass = "gr-qc",
	reportNumber = "IHES-P-09-28",
	doi = "10.1103/PhysRevD.80.044017",
	journal = "Phys. Rev. D",
	volume = "80",
	pages = "044017",
	year = "2009"
}

@article{Pani:2015hfa,
	author = "Pani, Paolo and Gualtieri, Leonardo and Maselli, Andrea and Ferrari, Valeria",
	title = "{Tidal deformations of a spinning compact object}",
	eprint = "1503.07365",
	archivePrefix = "arXiv",
	primaryClass = "gr-qc",
	doi = "10.1103/PhysRevD.92.024010",
	journal = "Phys. Rev. D",
	volume = "92",
	number = "2",
	pages = "024010",
	year = "2015"
}

@article{Pani:2015nua,
	author = "Pani, Paolo and Gualtieri, Leonardo and Ferrari, Valeria",
	title = "{Tidal Love numbers of a slowly spinning neutron star}",
	eprint = "1509.02171",
	archivePrefix = "arXiv",
	primaryClass = "gr-qc",
	doi = "10.1103/PhysRevD.92.124003",
	journal = "Phys. Rev. D",
	volume = "92",
	number = "12",
	pages = "124003",
	year = "2015"
}

@article{Chia:2020yla,
    author = "Chia, Horng Sheng",
    title = "{Tidal deformation and dissipation of rotating black holes}",
    eprint = "2010.07300",
    archivePrefix = "arXiv",
    primaryClass = "gr-qc",
    doi = "10.1103/PhysRevD.104.024013",
    journal = "Phys. Rev. D",
    volume = "104",
    number = "2",
    pages = "024013",
    year = "2021"
}

@article{Cardoso:2019vof,
    author = "Cardoso, Vitor and Gualtieri, Leonardo and Moore, Christopher J.",
    title = "{Gravitational waves and higher dimensions: Love numbers and Kaluza-Klein excitations}",
    eprint = "1910.09557",
    archivePrefix = "arXiv",
    primaryClass = "gr-qc",
    doi = "10.1103/PhysRevD.100.124037",
    journal = "Phys. Rev. D",
    volume = "100",
    number = "12",
    pages = "124037",
    year = "2019"
}

@article{Porto:2016zng,
	author = "Porto, Rafael A.",
	title = "{The Tune of Love and the Nature(ness) of Spacetime}",
	eprint = "1606.08895",
	archivePrefix = "arXiv",
	primaryClass = "gr-qc",
	doi = "10.1002/prop.201600064",
	journal = "Fortsch. Phys.",
	volume = "64",
	number = "10",
	pages = "723--729",
	year = "2016"
}

@article{LeTiec:2020spy,
	author = "Le Tiec, Alexandre and Casals, Marc",
	title = "{Spinning Black Holes Fall in Love}",
	eprint = "2007.00214",
	archivePrefix = "arXiv",
	primaryClass = "gr-qc",
	doi = "10.1103/PhysRevLett.126.131102",
	journal = "Phys. Rev. Lett.",
	volume = "126",
	number = "13",
	pages = "131102",
	year = "2021"
}

@article{Cardoso:2018ptl,
	author = "Cardoso, Vitor and Kimura, Masashi and Maselli, Andrea and Senatore, Leonardo",
	title = "{Black Holes in an Effective Field Theory Extension of General Relativity}",
	eprint = "1808.08962",
	archivePrefix = "arXiv",
	primaryClass = "gr-qc",
	doi = "10.1103/PhysRevLett.121.251105",
	journal = "Phys. Rev. Lett.",
	volume = "121",
	number = "25",
	pages = "251105",
	year = "2018"
}

@article{Gurlebeck:2015xpa,
	author = {G\"urlebeck, Norman},
	title = "{No-hair theorem for Black Holes in Astrophysical Environments}",
	eprint = "1503.03240",
	archivePrefix = "arXiv",
	primaryClass = "gr-qc",
	doi = "10.1103/PhysRevLett.114.151102",
	journal = "Phys. Rev. Lett.",
	volume = "114",
	number = "15",
	pages = "151102",
	year = "2015"
}

@article{Abdalla:2007dz,
    author = "Abdalla, E. and Konoplya, R. A. and Zhidenko, A.",
    title = "{Perturbations of Schwarzschild black holes in laboratories}",
    eprint = "0706.2489",
    archivePrefix = "arXiv",
    primaryClass = "hep-th",
    doi = "10.1088/0264-9381/24/23/012",
    journal = "Class. Quant. Grav.",
    volume = "24",
    pages = "5901--5910",
    year = "2007"
}

@article{Damour:2009vw,
    author = "Damour, Thibault and Nagar, Alessandro",
    title = "{Relativistic tidal properties of neutron stars}",
    eprint = "0906.0096",
    archivePrefix = "arXiv",
    primaryClass = "gr-qc",
    doi = "10.1103/PhysRevD.80.084035",
    journal = "Phys. Rev. D",
    volume = "80",
    pages = "084035",
    year = "2009"
}

@article{Ivanov:2022hlo,
    author = "Ivanov, Mikhail M. and Zhou, Zihan",
    title = "{Revisiting the matching of black hole tidal responses: A systematic study of relativistic and logarithmic corrections}",
    eprint = "2208.08459",
    archivePrefix = "arXiv",
    primaryClass = "hep-th",
    doi = "10.1103/PhysRevD.107.084030",
    journal = "Phys. Rev. D",
    volume = "107",
    number = "8",
    pages = "084030",
    year = "2023"
}

@article{Kol:2011vg,
    author = "Kol, Barak and Smolkin, Michael",
    title = "{Black hole stereotyping: Induced gravito-static polarization}",
    eprint = "1110.3764",
    archivePrefix = "arXiv",
    primaryClass = "hep-th",
    doi = "10.1007/JHEP02(2012)010",
    journal = "JHEP",
    volume = "02",
    pages = "010",
    year = "2012"
}

@article{LeTiec:2020bos,
	author = "Le Tiec, Alexandre and Casals, Marc and Franzin, Edgardo",
	title = "{Tidal Love Numbers of Kerr Black Holes}",
	eprint = "2010.15795",
	archivePrefix = "arXiv",
	primaryClass = "gr-qc",
	doi = "10.1103/PhysRevD.103.084021",
	journal = "Phys. Rev. D",
	volume = "103",
	number = "8",
	pages = "084021",
	year = "2021"
}

@article{Charalambous:2021kcz,
    author = "Charalambous, Panagiotis and Dubovsky, Sergei and Ivanov, Mikhail M.",
    title = "{Hidden Symmetry of Vanishing Love Numbers}",
    eprint = "2103.01234",
    archivePrefix = "arXiv",
    primaryClass = "hep-th",
    reportNumber = "INR-TH-2021-003",
    doi = "10.1103/PhysRevLett.127.101101",
    journal = "Phys. Rev. Lett.",
    volume = "127",
    number = "10",
    pages = "101101",
    year = "2021"
}

@article{Hui:2022vbh,
    author = "Hui, Lam and Joyce, Austin and Penco, Riccardo and Santoni, Luca and Solomon, Adam R.",
    title = "{Near-zone symmetries of Kerr black holes}",
    eprint = "2203.08832",
    archivePrefix = "arXiv",
    primaryClass = "hep-th",
    doi = "10.1007/JHEP09(2022)049",
    journal = "JHEP",
    volume = "09",
    pages = "049",
    year = "2022"
}

@article{Hui:2020xxx,
    author = "Hui, Lam and Joyce, Austin and Penco, Riccardo and Santoni, Luca and Solomon, Adam R.",
    title = "{Static response and Love numbers of Schwarzschild black holes}",
    eprint = "2010.00593",
    archivePrefix = "arXiv",
    primaryClass = "hep-th",
    doi = "10.1088/1475-7516/2021/04/052",
    journal = "JCAP",
    volume = "04",
    pages = "052",
    year = "2021"
}

@article{Maggiore:2019uih,
    author = "Maggiore, Michele and others",
    title = "{Science Case for the Einstein Telescope}",
    eprint = "1912.02622",
    archivePrefix = "arXiv",
    primaryClass = "astro-ph.CO",
    doi = "10.1088/1475-7516/2020/03/050",
    journal = "JCAP",
    volume = "03",
    pages = "050",
    year = "2020"
}

@article{Sathyaprakash:2019yqt,
    author = "Sathyaprakash, B. S. and others",
    title = "{Extreme Gravity and Fundamental Physics}",
    eprint = "1903.09221",
    archivePrefix = "arXiv",
    primaryClass = "astro-ph.HE",
    month = "3",
    year = "2019"
}

@article{Punturo:2010zz,
    author = "Punturo, M. and others",
    editor = "Ricci, Fulvio",
    title = "{The Einstein Telescope: A third-generation gravitational wave observatory}",
    doi = "10.1088/0264-9381/27/19/194002",
    journal = "Class. Quant. Grav.",
    volume = "27",
    pages = "194002",
    year = "2010"
}

@article{Kalogera:2021bya,
    author = "Kalogera, Vicky and others",
    title = "{The Next Generation Global Gravitational Wave Observatory: The Science Book}",
    eprint = "2111.06990",
    archivePrefix = "arXiv",
    primaryClass = "gr-qc",
    month = "11",
    year = "2021"
}

@article{Maselli:2018fay,
    author = "Maselli, Andrea and Pani, Paolo and Cardoso, Vitor and Abdelsalhin, Tiziano and Gualtieri, Leonardo and Ferrari, Valeria",
    title = "{From micro to macro and back: probing near-horizon quantum structures with gravitational waves}",
    eprint = "1811.03689",
    archivePrefix = "arXiv",
    primaryClass = "gr-qc",
    doi = "10.1088/1361-6382/ab30ff",
    journal = "Class. Quant. Grav.",
    volume = "36",
    number = "16",
    pages = "167001",
    year = "2019"
}

@article{Datta:2021hvm,
    author = "Datta, Sayak",
    title = "{Probing horizon scale quantum effects with Love}",
    eprint = "2107.07258",
    archivePrefix = "arXiv",
    primaryClass = "gr-qc",
    reportNumber = "LIGO-P210025",
    doi = "10.1088/1361-6382/ac9ae4",
    journal = "Class. Quant. Grav.",
    volume = "39",
    number = "22",
    pages = "225016",
    year = "2022"
}

@article{Sharma:2024hlz,
    author = "Sharma, Chanchal and Ghosh, Rajes and Sarkar, Sudipta",
    title = "{Exploring ladder symmetry and Love numbers for static and rotating black holes}",
    eprint = "2401.00703",
    archivePrefix = "arXiv",
    primaryClass = "gr-qc",
    doi = "10.1103/PhysRevD.109.L041505",
    journal = "Phys. Rev. D",
    volume = "109",
    number = "4",
    pages = "L041505",
    year = "2024"
}

@article{Branchesi:2023mws,
    author = "Branchesi, Marica and others",
    title = "{Science with the Einstein Telescope: a comparison of different designs}",
    eprint = "2303.15923",
    archivePrefix = "arXiv",
    primaryClass = "gr-qc",
    reportNumber = "ET-0084A-23",
    doi = "10.1088/1475-7516/2023/07/068",
    journal = "JCAP",
    volume = "07",
    pages = "068",
    year = "2023"
}

@article{Pani:2015tga,
    author = "Pani, Paolo",
    title = "{I-Love-Q relations for gravastars and the approach to the black-hole limit}",
    eprint = "1506.06050",
    archivePrefix = "arXiv",
    primaryClass = "gr-qc",
    doi = "10.1103/PhysRevD.95.049902",
    journal = "Phys. Rev. D",
    volume = "92",
    number = "12",
    pages = "124030",
    year = "2015",
    note = "[Erratum: Phys.Rev.D 95, 049902 (2017)]"
}

@article{Hinderer:2007mb,
    author = "Hinderer, Tanja",
    title = "{Tidal Love numbers of neutron stars}",
    eprint = "0711.2420",
    archivePrefix = "arXiv",
    primaryClass = "astro-ph",
    doi = "10.1086/533487",
    journal = "Astrophys. J.",
    volume = "677",
    pages = "1216--1220",
    year = "2008",
    note = "[Erratum: Astrophys.J. 697, 964 (2009)]"
}

@article{Charalambous:2021mea,
	title        = {{On the Vanishing of Love Numbers for Kerr Black Holes}},
	author       = {Charalambous, Panagiotis and Dubovsky, Sergei and Ivanov, Mikhail M.},
	year         = 2021,
	journal      = {JHEP},
	volume       = {05},
	pages        = {038},
	doi          = {10.1007/JHEP05(2021)038},
	eprint       = {2102.08917},
	archiveprefix = {arXiv},
	primaryclass = {hep-th},
	reportnumber = {INR-TH-2021-001}
}

@article{Binnington:2009bb,
	title        = {{Relativistic theory of tidal Love numbers}},
	author       = {Binnington, Taylor and Poisson, Eric},
	year         = 2009,
	journal      = {Phys. Rev. D},
	volume       = 80,
	pages        = {084018},
	doi          = {10.1103/PhysRevD.80.084018},
	eprint       = {0906.1366},
	archiveprefix = {arXiv},
	primaryclass = {gr-qc}
}

@article{Cardoso:2017cfl,
	title        = {{Testing strong-field gravity with tidal Love numbers}},
	author       = {Cardoso, Vitor and Franzin, Edgardo and Maselli, Andrea and Pani, Paolo and Raposo, Guilherme},
	year         = 2017,
	journal      = {Phys. Rev. D},
	volume       = 95,
	number       = 8,
	pages        = {084014},
	doi          = {10.1103/PhysRevD.95.084014},
	note         = {[Addendum: Phys.Rev.D 95, 089901 (2017)]},
	eprint       = {1701.01116},
	archiveprefix = {arXiv},
	primaryclass = {gr-qc}
}

@article{Charalambous:2024tdj,
    author = "Charalambous, Panagiotis",
    title = "{Love numbers and Love symmetries for p-form and gravitational perturbations of higher-dimensional spherically symmetric black holes}",
    eprint = "2402.07574",
    archivePrefix = "arXiv",
    primaryClass = "hep-th",
    doi = "10.1007/JHEP04(2024)122",
    journal = "JHEP",
    volume = "04",
    pages = "122",
    year = "2024"
}

@article{Charalambous:2023jgq,
    author = "Charalambous, Panagiotis and Ivanov, Mikhail M.",
    title = "{Scalar Love numbers and Love symmetries of 5-dimensional Myers-Perry black holes}",
    eprint = "2303.16036",
    archivePrefix = "arXiv",
    primaryClass = "hep-th",
    doi = "10.1007/JHEP07(2023)222",
    journal = "JHEP",
    volume = "07",
    pages = "222",
    year = "2023"
}

@article{Rai:2024lho,
    author = "Rai, Mudit and Santoni, Luca",
    title = {{Ladder symmetries and Love numbers of Reissner-Nordstr\"om black holes}},
    eprint = "2404.06544",
    archivePrefix = "arXiv",
    primaryClass = "gr-qc",
    doi = "10.1007/JHEP07(2024)098",
    journal = "JHEP",
    volume = "07",
    pages = "098",
    year = "2024"
}

@article{GuerraChaves:2019foa,
    author = "Guerra Chaves, Andreas and Hinderer, Tanja",
    title = "{Probing the equation of state of neutron star matter with gravitational waves from binary inspirals in light of GW170817: a brief review}",
    eprint = "1912.01461",
    archivePrefix = "arXiv",
    primaryClass = "nucl-th",
    doi = "10.1088/1361-6471/ab45be",
    journal = "J. Phys. G",
    volume = "46",
    number = "12",
    pages = "123002",
    year = "2019"
}

@article{Chatziioannou:2020pqz,
    author = "Chatziioannou, Katerina",
    title = "{Neutron star tidal deformability and equation of state constraints}",
    eprint = "2006.03168",
    archivePrefix = "arXiv",
    primaryClass = "gr-qc",
    doi = "10.1007/s10714-020-02754-3",
    journal = "Gen. Rel. Grav.",
    volume = "52",
    number = "11",
    pages = "109",
    year = "2020"
}

@article{LISA:2017pwj,
    author = "Amaro-Seoane, Pau and others",
    collaboration = "LISA",
    title = "{Laser Interferometer Space Antenna}",
    eprint = "1702.00786",
    archivePrefix = "arXiv",
    primaryClass = "astro-ph.IM",
    month = "2",
    year = "2017"
}

@article{Cardoso:2021wlq,
    author = "Cardoso, Vitor and Destounis, Kyriakos and Duque, Francisco and Macedo, Rodrigo Panosso and Maselli, Andrea",
    title = "{Black holes in galaxies: Environmental impact on gravitational-wave generation and propagation}",
    eprint = "2109.00005",
    archivePrefix = "arXiv",
    primaryClass = "gr-qc",
    doi = "10.1103/PhysRevD.105.L061501",
    journal = "Phys. Rev. D",
    volume = "105",
    number = "6",
    pages = "L061501",
    year = "2022"
}

@ARTICLE{1909MNRAS..69..476L,
       author = {{Love}, A.~E.~H.},
        title = "{Earth, the yielding of the, to disturbing forces}",
      journal = {Mont. Not. Roy. Astr. Soc.},
         year = 1909,
        month = apr,
       volume = {69},
        pages = {476},
          doi = {10.1093/mnras/69.6.476},
       adsurl = {https://ui.adsabs.harvard.edu/abs/1909MNRAS..69..476L}
}

@article{Kosmopoulos:2025rfj,
    author = "Kosmopoulos, Dimitrios and Perrone, Davide and Solon, Mikhail",
    title = "{Dynamical Love Numbers for Black Holes and Beyond from Shell Effective Field Theory}",
    eprint = "2512.04002",
    archivePrefix = "arXiv",
    primaryClass = "hep-th",
    month = "12",
    year = "2025"
}

@article{DeLuca:2023mio,
    author = "De Luca, Valerio and Khoury, Justin and Wong, Sam S. C.",
    title = "{Nonlinearities in the tidal Love numbers of black holes}",
    eprint = "2305.14444",
    archivePrefix = "arXiv",
    primaryClass = "gr-qc",
    doi = "10.1103/PhysRevD.108.024048",
    journal = "Phys. Rev. D",
    volume = "108",
    number = "2",
    pages = "024048",
    year = "2023"
}

@article{Saketh:2023bul,
    author = "Saketh, M. V. S. and Zhou, Zihan and Ivanov, Mikhail M.",
    title = "{Dynamical tidal response of Kerr black holes from scattering amplitudes}",
    eprint = "2307.10391",
    archivePrefix = "arXiv",
    primaryClass = "hep-th",
    doi = "10.1103/PhysRevD.109.064058",
    journal = "Phys. Rev. D",
    volume = "109",
    number = "6",
    pages = "064058",
    year = "2024"
}

@article{Bhatt:2023zsy,
    author = "Bhatt, Rajendra Prasad and Chakraborty, Sumanta and Bose, Sukanta",
    title = "{Addressing issues in defining the Love numbers for black holes}",
    eprint = "2306.13627",
    archivePrefix = "arXiv",
    primaryClass = "gr-qc",
    reportNumber = "LIGO-P2300180",
    doi = "10.1103/PhysRevD.108.084013",
    journal = "Phys. Rev. D",
    volume = "108",
    number = "8",
    pages = "084013",
    year = "2023"
}

@article{Barcelo:2005fc,
    author = "Barcelo, Carlos and Liberati, Stefano and Visser, Matt",
    title = "{Analogue gravity}",
    eprint = "gr-qc/0505065",
    archivePrefix = "arXiv",
    doi = "10.12942/lrr-2005-12",
    journal = "Living Rev. Rel.",
    volume = "8",
    pages = "12",
    year = "2005"
}

@article{Berti:2004ju,
    author = "Berti, Emanuele and Cardoso, Vitor and Lemos, Jose P. S.",
    title = "{Quasinormal modes and classical wave propagation in analogue black holes}",
    eprint = "gr-qc/0408099",
    archivePrefix = "arXiv",
    doi = "10.1103/PhysRevD.70.124006",
    journal = "Phys. Rev. D",
    volume = "70",
    pages = "124006",
    year = "2004"
}

@article{Cardoso:2004fi,
    author = "Cardoso, Vitor and Lemos, Jose P. S. and Yoshida, Shijun",
    title = "{Quasinormal modes and stability of the rotating acoustic black hole: Numerical analysis}",
    eprint = "gr-qc/0410107",
    archivePrefix = "arXiv",
    doi = "10.1103/PhysRevD.70.124032",
    journal = "Phys. Rev. D",
    volume = "70",
    pages = "124032",
    year = "2004"
}

@inproceedings{Cardoso:2005ij,
    author = "Cardoso, Vitor",
    title = "{Acoustic black holes}",
    booktitle = "{5th International Workshop on New Worlds in Astroparticle Physics}",
    eprint = "physics/0503042",
    archivePrefix = "arXiv",
    doi = "10.1142/9789812774439_0026",
    pages = "245--251",
    month = "3",
    year = "2005"
}

@article{Kim:2004sf,
    author = "Kim, Sung-Won and Kim, Won Tae and Oh, John J.",
    title = "{Decay rate and low energy near horizon dynamics of acoustic black holes}",
    eprint = "gr-qc/0409003",
    archivePrefix = "arXiv",
    doi = "10.1016/j.physletb.2005.01.012",
    journal = "Phys. Lett. B",
    volume = "608",
    pages = "10--16",
    year = "2005"
}

@article{Lepe:2004kv,
    author = "Lepe, Samuel and Saavedra, Joel",
    title = "{Quasinormal modes, superradiance and area spectrum for 2+1 acoustic black holes}",
    eprint = "gr-qc/0410074",
    archivePrefix = "arXiv",
    reportNumber = "GACG-04-13",
    doi = "10.1016/j.physletb.2005.05.021",
    journal = "Phys. Lett. B",
    volume = "617",
    pages = "174--181",
    year = "2005"
}

@article{Saavedra:2005ug,
    author = "Saavedra, Joel",
    title = "{Quasinormal modes of Unruh's acoustic black hole}",
    eprint = "gr-qc/0508040",
    archivePrefix = "arXiv",
    doi = "10.1142/S0217732306019712",
    journal = "Mod. Phys. Lett. A",
    volume = "21",
    pages = "1601--1608",
    year = "2006"
}

@article{Visser:1997ux,
    author = "Visser, Matt",
    title = "{Acoustic black holes: Horizons, ergospheres, and Hawking radiation}",
    eprint = "gr-qc/9712010",
    archivePrefix = "arXiv",
    doi = "10.1088/0264-9381/15/6/024",
    journal = "Class. Quant. Grav.",
    volume = "15",
    pages = "1767--1791",
    year = "1998"
}

@article{Katagiri:2024fpn,
    author = "Katagiri, Takuya and Cardoso, Vitor and Ikeda, Tact and Yagi, Kent",
    title = "{Tidal response beyond vacuum general relativity with a canonical definition}",
    eprint = "2410.02531",
    archivePrefix = "arXiv",
    primaryClass = "gr-qc",
    reportNumber = "RUP-24-19",
    doi = "10.1103/PhysRevD.111.084081",
    journal = "Phys. Rev. D",
    volume = "111",
    number = "8",
    pages = "084081",
    year = "2025"
}

@article{Singh:2024qfw,
    author = "Singh, Balbeer and Padhi, Nibedita and Nayak, Rashmi R.",
    title = "{Circular orbits and chaos bound in slow-rotating curved acoustic black holes}",
    eprint = "2405.12337",
    archivePrefix = "arXiv",
    primaryClass = "hep-th",
    doi = "10.1140/epjc/s10052-025-14311-w",
    journal = "Eur. Phys. J. C",
    volume = "85",
    number = "5",
    pages = "570",
    year = "2025"
}

@article{Katagiri:2024wbg,
    author = "Katagiri, Takuya and Yagi, Kent and Cardoso, Vitor",
    title = "{Relativistic dynamical tides: Subtleties and calibration}",
    eprint = "2409.18034",
    archivePrefix = "arXiv",
    primaryClass = "gr-qc",
    doi = "10.1103/PhysRevD.111.084080",
    journal = "Phys. Rev. D",
    volume = "111",
    number = "8",
    pages = "084080",
    year = "2025"
}

@article{Iteanu:2024dvx,
    author = "Iteanu, Simon and Riva, Massimiliano Maria and Santoni, Luca and Savi\'c, Nikola and Vernizzi, Filippo",
    title = "{Vanishing of quadratic Love numbers of Schwarzschild black holes}",
    eprint = "2410.03542",
    archivePrefix = "arXiv",
    primaryClass = "gr-qc",
    reportNumber = "DESY 24-141",
    doi = "10.1007/JHEP02(2025)174",
    journal = "JHEP",
    volume = "02",
    pages = "174",
    year = "2025"
}

@article{Kehagias:2024rtz,
    author = "Kehagias, Alex and Riotto, Antonio",
    title = "{Black holes in a gravitational field: the non-linear static love number of Schwarzschild black holes vanishes}",
    eprint = "2410.11014",
    archivePrefix = "arXiv",
    primaryClass = "gr-qc",
    doi = "10.1088/1475-7516/2025/05/039",
    journal = "JCAP",
    volume = "05",
    pages = "039",
    year = "2025"
}

\end{document}